\documentclass[fleqn,usenatbib]{mnras}

\usepackage[T1]{fontenc}

\DeclareRobustCommand{\VAN}[3]{#2}
\let\VANthebibliography\thebibliography
\def\thebibliography{\DeclareRobustCommand{\VAN}[3]{##3}\VANthebibliography}

\usepackage{graphicx}	
\usepackage{amsmath}	
\usepackage{amssymb}	

\usepackage{newtxtext,newtxmath}

\newcommand{\radm}{ rad m$^{-2}$ }
\newcommand{\pccm}{ pc cm$^{-3}$ }

\title[RM variations in GC pulsars]{Rotation measure variations in Galactic Centre pulsars}

\author[F. Abbate et al.]{
F. Abbate,$^{1}$\thanks{E-mail: abbate@mpifr-bonn.mpg.de}
A. Noutsos,$^{2}$
G. Desvignes,$^{1}$
R.~S.~Wharton,$^{3}$
P. Torne,$^{4,1}$
M. Kramer,$^{1,5}$
R.~P.~Eatough,$^{6,1}$
\and
R. Karuppusamy,$^{1}$
K. Liu,$^{1}$
L. Shao$^{7,1,6}$
and J. Wongphechauxsorn$^{1}$
\\
$^{1}$Max-Planck-Institut f\"{u}r Radioastronomie, Auf dem H\"{u}gel 69, D-53121 Bonn, Germany\\
$^{2}$ SKA Observatory, Jodrell Bank, Lower Withington, Macclesfield SK11 9FT, UK \\
$^{3}$ Jet Propulsion Laboratory, California Institute of Technology, Pasadena, CA 91109, USA \\
$^{4}$Institut de Radioastronomie Millim\'etrique, Avda. Divina Pastora 7, Local 20, 18012 Granada, Spain\\
$^{5}$Jodrell Bank Centre for Astrophysics, Department of Physics and Astronomy, The University of Manchester, Manchester M13 9PL, UK\\
$^{6}$National Astronomical Observatories, Chinese Academy of Sciences, 20A Datun Road, Chaoyang District, Beijing 100101, P. R. China\\
$^{7}$Kavli Institute for Astronomy and Astrophysics, Peking University, Beijing 100871, People's Republic of China
}

\date{Accepted XXX. Received YYY; in original form ZZZ}

\pubyear{2023}

\begin{document}
\label{firstpage}
\pagerange{\pageref{firstpage}--\pageref{lastpage}}
\maketitle

\begin{abstract}
We report the results of an observational campaign using the Effelsberg 100-m telescope of the pulsars J1746$-$2849, J1746$-$2850, J1746$-$2856 and J1745$-$2912 located in the Central Molecular Zone (CMZ) close to the Galactic centre in order to study rotation measure (RM) variations. We report for the first time the RM value of PSR J1746$-$2850 to be $-12234 \pm 181$ rad m$^{-2}$. This pulsar shows significant variations of RM of $300-400$ rad m$^{-2}$ over the course of months to years that suggest a strongly magnetized environment. The structure function analysis of the RM of PSR J1746$-$2850 revealed a steep power-law index of $1.87_{-0.3}^{+0.4}$ comparable to the value expected for isotropic turbulence.
This pulsar also showed large dispersion measure (DM) variation of  $\sim 50$ pc cm$^{-3}$ in an event lasting a few months where the RM increased by $\sim 200$ rad m$^{-2}$. The large difference in RM between PSR J1746$-$2849 and PSR J1746$-$2850 despite the small angular separation reveals the presence of a magnetic field of at least 70 $\mu$G in the CMZ and can explain the lack of polarization in the radio images of the region. These results contribute to our understanding of the magnetic field in the CMZ and show similarities between the RM behaviours of these pulsars and some fast radio bursts (FRBs).

\end{abstract}




\section{Introduction}

The interstellar medium (ISM) surrounding the supermassive black hole Sagittarius A$^*$ (Sgr A$^*$) at the centre of the Milky way is quite extreme when compared to the Galactic disk. The region surrounding Sgr A$^*$ with a radius of $\sim 150$ pc, called the central molecular zone (CMZ, \citealt{Morris1996}), has densities \citep{Martin2004, Ferriere2007} and cosmic-ray energy density \citep{Oka2019} 2$-$3 orders of magnitude larger than the rest of the Galaxy. The situation is similar for the magnetic fields. While direct measurements of the magnetic fields in the region do not agree on a single value \citep{Ferriere2009}, arguments regarding the non-thermal filaments (NTFs), cosmic-ray density and turbulent energy suggest that interstellar magnetic fields range between 100 $\mu$G and 1 mG \citep{Ferriere2009, Oka2019}. This is about two orders of magnitude higher than the magnetic field in the Galactic disk \citep{Han2006,Orlando2013}.

One technique that can be used to probe the magnetic field is through the rotation measure (RM) and dispersion measure (DM) of the pulsars located within this region. There are 6 known pulsars located within the CMZ, PSR J1745$-$2912, PSR J1746$-$2856, PSR J1746$-$2849, PSR J1746$-$2850, PSR J1745$-$2910 \citep{Johnston2006,Deneva2009} and the Galactic centre magnetar PSR J1745$-$2900 \citep{Kennea2013,Mori2013,eatough2013}. These pulsars have among the highest DMs and RMs of any known pulsar \citep{Schnitzeler2016}, which reinforces the idea that high densities and strong magnetic fields permeate the CMZ region.

The RM of PSR J1745$-$2900, located just 3 arcseconds away from Sgr A$^*$, has been closely monitored throughout the years since its discovery and shows very strong variations with an increase of $\sim 3500$ \radm from $-66,960$ \radm to $-63,402$ \radm in the span of 3 years \citep{Desvignes2018}. This difference has been attributed to the rapidly changing magnetic environment close to Sgr A$^*$. Such strong variability is rare for pulsars and is only seen in another class of objects, the Fast Radio Bursts (FRBs, e.g. \citealt{Petroff2019,Cordes2019}).
In this case, the RM variability over a few months or years ranges from $\sim 50$\radm for FRB 20180916B \citep{Mckinven2022} to a few tens of thousands of \radm for FRB 121102 \citep{Hilmarsson2021} and FRB 20190520B \citep{Anna-Thomas2023}. The similar variations over comparable timescales suggest that the surrounding environments might have similar levels of density and magnetic fields.

\begin{figure*}
\centering
	\includegraphics[width=\textwidth]{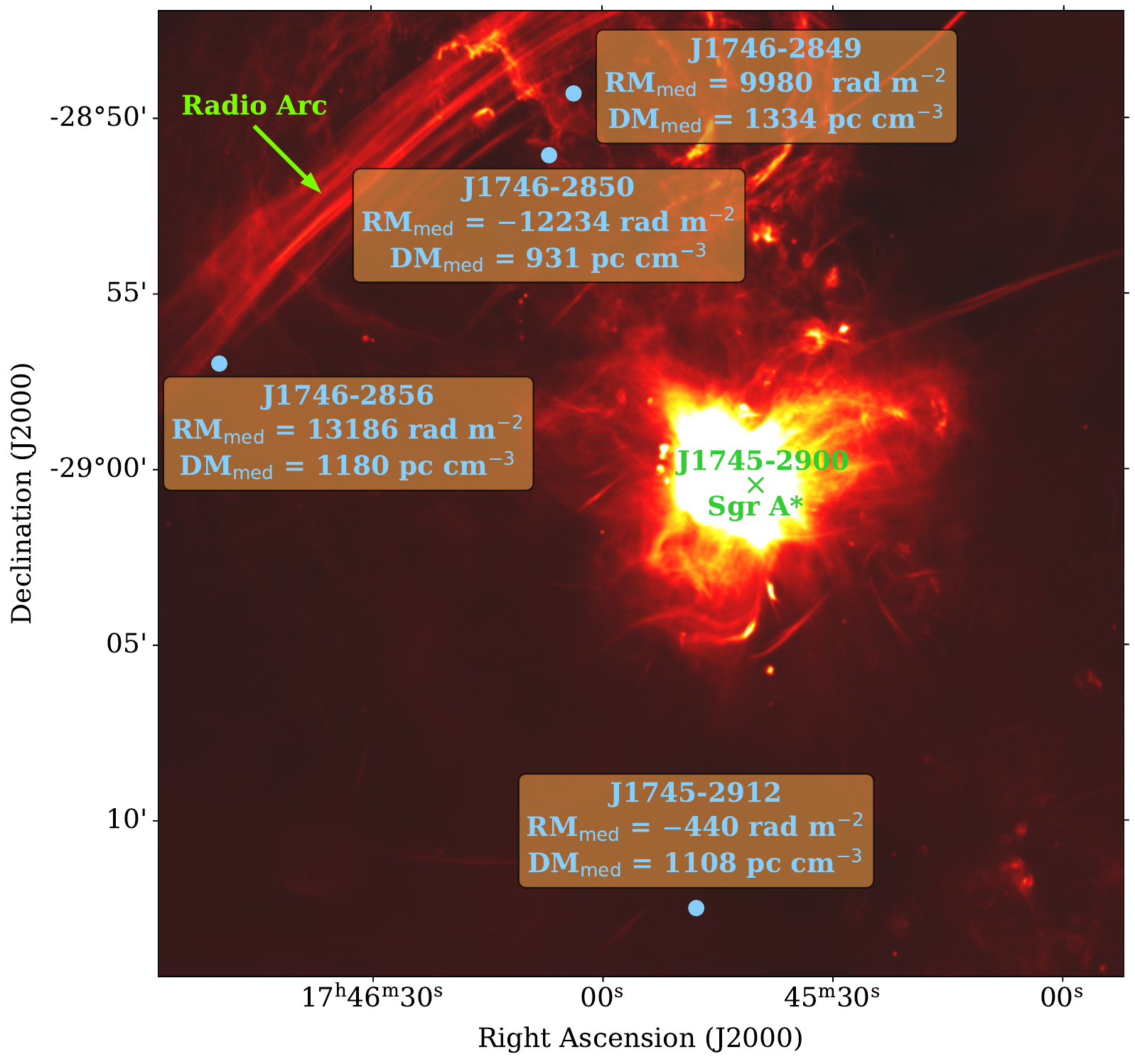}
    \caption{Image of the Galactic centre region showing in blue the positions of the four pulsars observed. The background comes from L-band observations obtained at MeerKAT \citep{Heywood2022}. For each pulsar we report the median values of RM and DM measured in the current work. The position of Sgr A$^*$, the magnetar PSR J1745$-$2900 and the Radio Arc NTF are shown in green. At the distance of Sgr A$^*$, the physical size of the image is 60x60 pc. 
    }
  	\label{fig:pulsars_positions}
\end{figure*}

In this paper we show the results of a 3 year long observational campaign on PSR J1746$-$2849, PSR J1746$-$2850, PSR J1746$-$2856 and PSR J1745$-$2912 in order to determine the variability of RM over time. Studying the extent of this variability could help probe the properties of the magnetic field in the CMZ compared to the surrounding of Sgr A$^*$. Additionally, the RM variability of these pulsars will provide important clues for the interpretation of variability seen in FRBs.

\section{Observations}

The observations were carried out with the Effelsberg 100-m radio telescope of the Max Planck Institute for Radio Astronomy using the S45 broadband receiver. This receiver has two 2 GHz bands (between 4 and 8 GHz) that are fed into the PSRIX2 backend, consisting of two CASPER\footnote{\url{https://casper.berkeley.edu/}} ROACH2 boards. The signal is digitized creating a total of 4096 frequency channels sampled every 131 $\mu$s and recorded in full-Stokes. 

PSR J1745-2910 was not detected in our first observation probably due to its high variability and/or dimming over time. 
It is important to note that this pulsar has only been observed at the Green Bank Telescope at a frequency of 2 GHz. The uncertainty of the position, given by the beam of $\sim$ 5 arcmin, is larger than the beam at the Effelsberg 100-m telescope at 6 GHz ($\sim$ 2 arcmin) meaning that the pulsar might be outside of the Effelsberg 100-m beam. 
For this reason we focused on PSR J1746$-$2849, PSR J1746$-$2850, PSR J1746$-$2856 and PSR J1745$-$2912. The positions of the four observed pulsars are shown in Fig. \ref{fig:pulsars_positions}. The observations were carried out from March 2019 to August 2022. The spacing of the observations is not uniform, and the minimum temporal difference between the observations is 10 d.

PSR J1746$-$2850 was the subject of a recent re-brightening following a few years of non-detections \citep{Dexter2017}. For this reason this pulsar was the target of a larger number of observations compared to the other pulsars. 
Phase-binned imaging observations with the VLA have determined that the position of 
PSR J1746$-$2850 is at R.A. (J2000) $17^h46^m06^s.959 \pm 0.002$ and DEC (J2000) $-28^{\circ}51'04''.54 \pm 0.08$
(Wharton et al., \emph{in prep}), which is about 25 arcseconds away from the timing position 
presented in \cite{Deneva2009}.  Throughout the paper we will use the imaging position.

\begin{figure*}
\centering
	\includegraphics[width=0.46\textwidth]{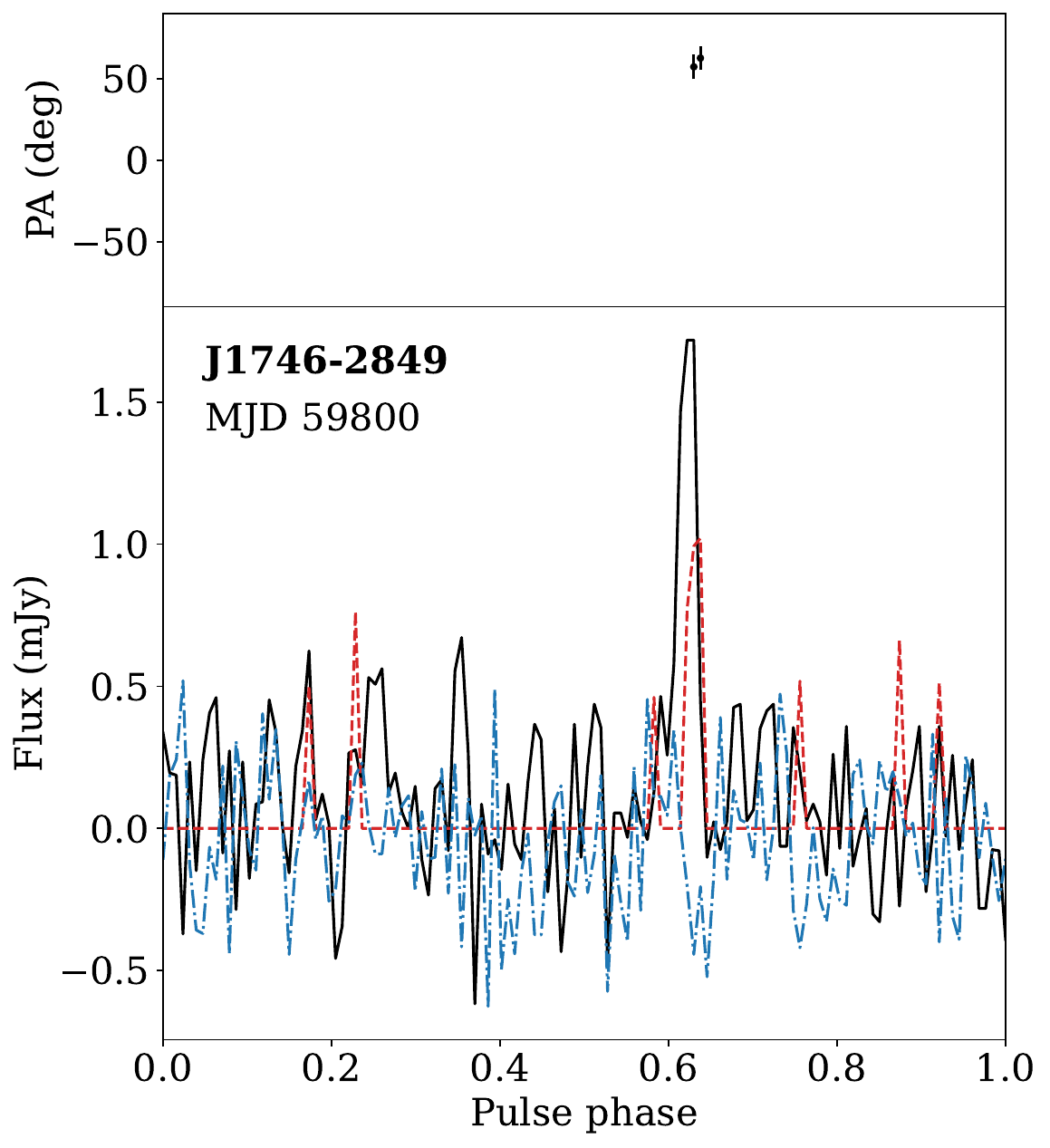}
	\,
	\includegraphics[width=0.46\textwidth]{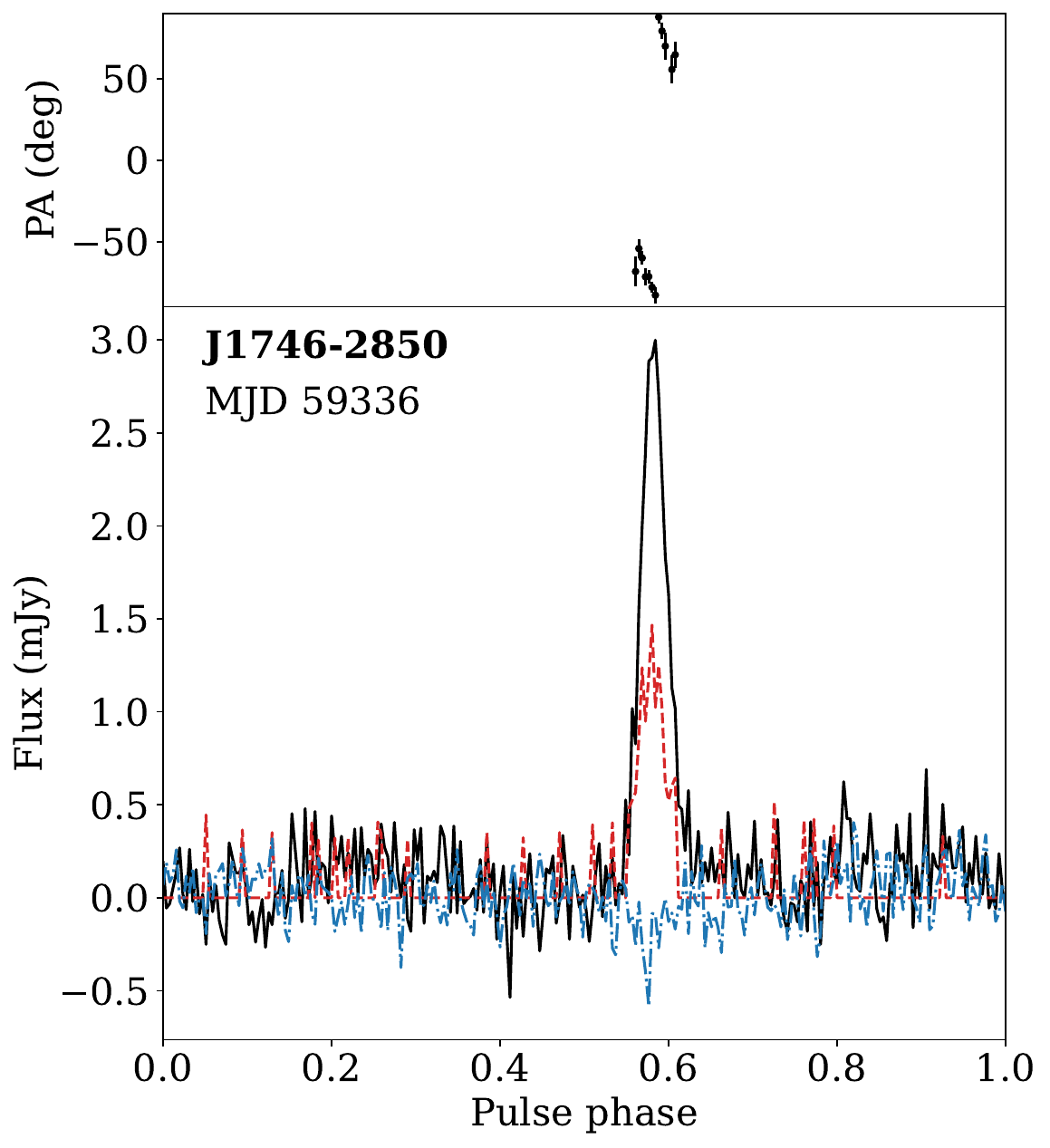}
	\,
	\includegraphics[width=0.46\textwidth]{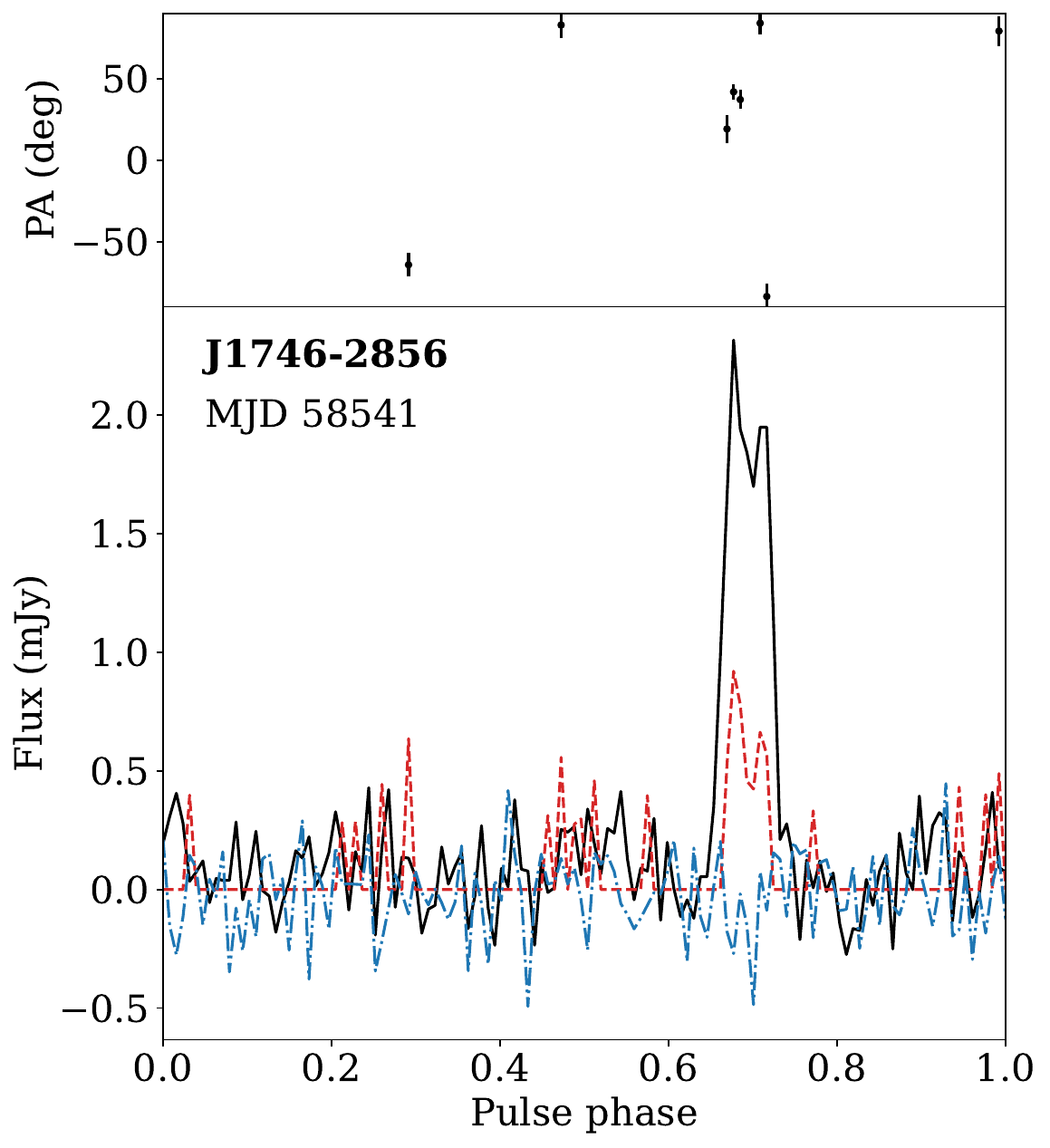}
	\,
	\includegraphics[width=0.46\textwidth]{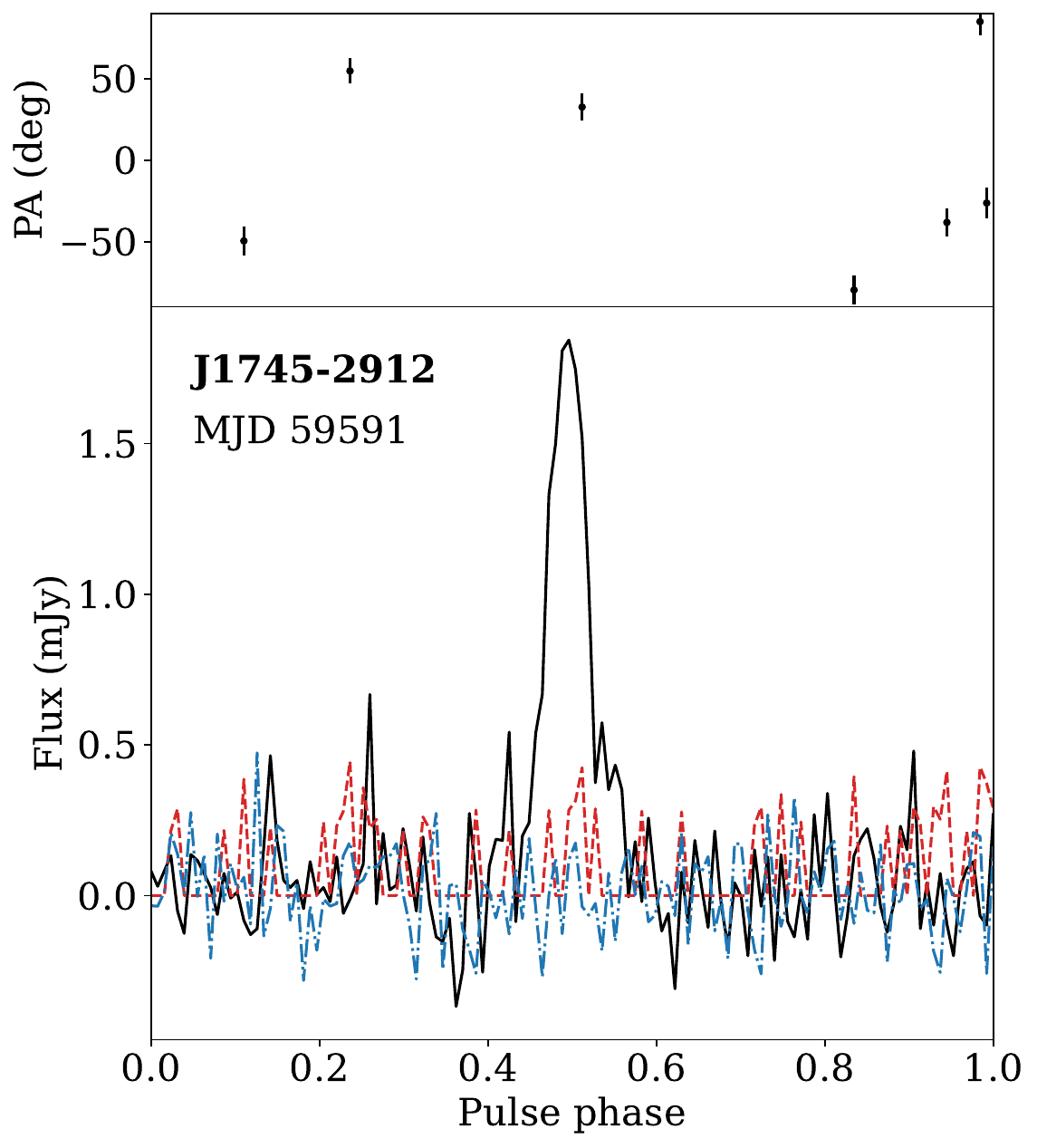}
  	\caption{Polarization profiles corrected for the best value of RM of one observation for each of the pulsars. The date corresponding to the observation is reported below the pulsar name in Modified Julian Date (MJD). The solid black line is the total intensity profile, the dashed red line is the linear polarization and the dot-dashed blue line is the circular polarization. The top panel shows the position angle (PA) at infinite frequency. The PA is shown only for the bins where the linear polarization has a S/N higher than 3.}
  	\label{fig:profiles}
\end{figure*}

\begin{table*}
\centering 
\caption{ Values of DM and RM for all of the observations of the observed pulsars. The errors in brackets are the 1$\sigma$ interval. In some cases the RM is not reported because the significance of the detection is lower than 4. See the appendix \ref{rm_fits} for details on the RM fit. }
\label{tab:results}
\footnotesize
\renewcommand{\arraystretch}{1}
\vskip 0.1cm
\begin{tabular}{c|ccc}
\hline
Pulsar name  & Date (MJD) & \multicolumn{1}{c}{DM} & \multicolumn{1}{c}{RM}\\
 & &  \multicolumn{1}{c}{(pc cm$^{-3}$)} &  \multicolumn{1}{c}{(rad m$^{-2}$)}\\
\hline
PSR J1746$-$2849 & 57148$^{*}$ &  & 101014(104)$^{*}$\\
           & 58544 & 1405(35) & 9980(160)\\
           & 58557 & 1326(21) & 9800(160)\\
           & 58623 & 1340(12) &  \\
           & 58865 & 1332(45) & 10050(260) \\
           & 59400 & 1334(30) & 9920(146)\\
           & 59480 & 1283(72) &  \\
           & 59800 & 1360(54) & 10122(160) \\
\hline           
PSR J1746$-$2850 & 58544 & 931(24) & -12363(44) \\
           & 58557 & 897(15) & -12533(86) \\
           & 58591 & 929(14) & -12507(33) \\
           & 58600 & 896(26) & -12453(68) \\
           & 58610 & 910(28) & -12529(46) \\
           & 58623 & 940(30) & -12510(60) \\
           & 58764 & 945(24) & -12340(120) \\
           & 58817 & 931(15) & -12335(50)  \\
           & 58831 & 952(40) & -12130(80) \\
           & 58865 & 886(19) & -12504(64) \\
           & 59314 & 952(18) & -12185(34) \\
           & 59336 & 916(11) & -12192(29) \\
           & 59357 & 940(11) & -12006(31) \\
           & 59377 & 956(19) & -12368(43) \\
           & 59437 & 957(6 ) & -12186(32)  \\
           & 59473 & 944(11) & -12199(23) \\
           & 59502 & 918(11) & -12240(33) \\
           & 59545 & 891(14) & -12228(40) \\
           & 59577 & 933(14) & -12166(29) \\
           & 59608 & 892(11) & -12167(34) \\
           & 59629 & 915(16) & -12116(29) \\
           & 59800 & 941(15) & -12034(100) \\
\hline
PSR J1746$-$2856 & 57148$^{*}$ &  & 13253(53)$^{*}$\\
           & 58541 & 1180(27) & 13186(47)\\
           & 58610 & 1231(26) & 13223(50)\\
           & 58831 & 1155(20) & 13096(100)\\
\hline
PSR J1745$-$2912 & 57148$^{*}$ &  & -535(107)$^{*}$\\
           & 58541 & 1110(4) & \\
           & 58591 & 1114(4) & -400(150)\\
           & 58801 & 1106(13) & -480(146)\\
           & 58817 & 1106(3) & -648(144)\\
\hline
$^{*}$ RM values measured in \cite{Schnitzeler2016}.
\end{tabular}
\end{table*}

\section{Data analysis} \label{sec:analysis}
The data were analyzed using standard \texttt{PSRCHIVE}\footnote{\url{http://psrchive.sourceforge.net}} \citep{Hotan2004,vanStraten2012} packages and calibrated in flux and polarization. The DMs were measured using \texttt{TEMPO2}\footnote{\url{https://bitbucket.org/psrsoft/tempo2/}} \citep{Hobbs2006} by dividing each observation in 16 frequency channels and extracting time of arrivals (ToAs) for each channel. Temporal variations of the scattering properties of the pulsars could mimic variations of DM. We tried looking for evidence of variations of the scattering using the software \texttt{PulsePortraiture} \citep{Pennucci2016,Pennucci2018} but no significant variation was detected.

The RMs were measured using a two-step approach similar to the one used by \cite{Han2018} and \cite{Johnston2020}. First we looked for the value of RM that maximises the linear polarization of the pulsed signal. We searched in a range of RMs between -50,000 and 50,000 rad m$^{-2}$ with a step size of 10 rad m$^{-2}$. After a preliminary value is found this way, we corrected the data for this value of RM, divided the data in 16 frequency channels and performed a fit of the position angle (PA) across the frequency band according to the formula: 
\begin{equation} \label{eq:PAfit}
    \Psi(\lambda) = {\rm RM}\, \lambda^2 + \Psi_0
\end{equation}
where $\lambda$ is the wavelength and $\Psi_0$ is the value of the PA as it was emitted by the pulsar. The measure of the value and error of the PA for each frequency channel follows the prescriptions described in \cite{Noutsos2008,Tiburzi2013,Abbate2020}. Following the prescriptions described in \cite{Sobey2019}, we discard the detection if the significance of peak in linear polarization is smaller than 4 and we multiply the error by two if the significance is between 4 and 8.
Examples, details of the fitting procedure and comparison with a method based on RM synthesis \citep{Brentjens2005} for each of the pulsars are shown in appendix \ref{rm_fits}.

\section{Results}

\begin{figure*}
\centering
	\includegraphics[width=0.48\textwidth]{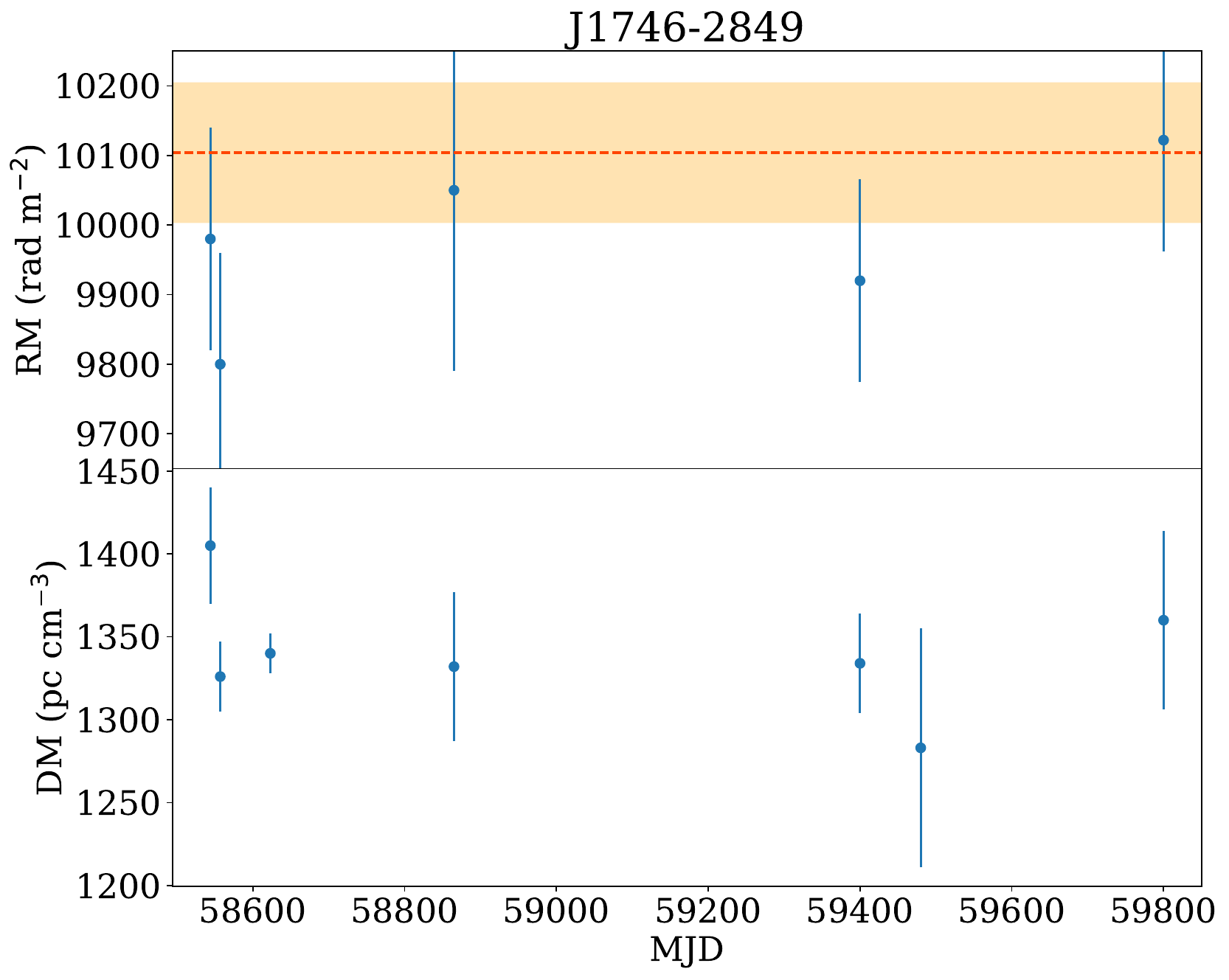}
	\,
	\includegraphics[width=0.48\textwidth]{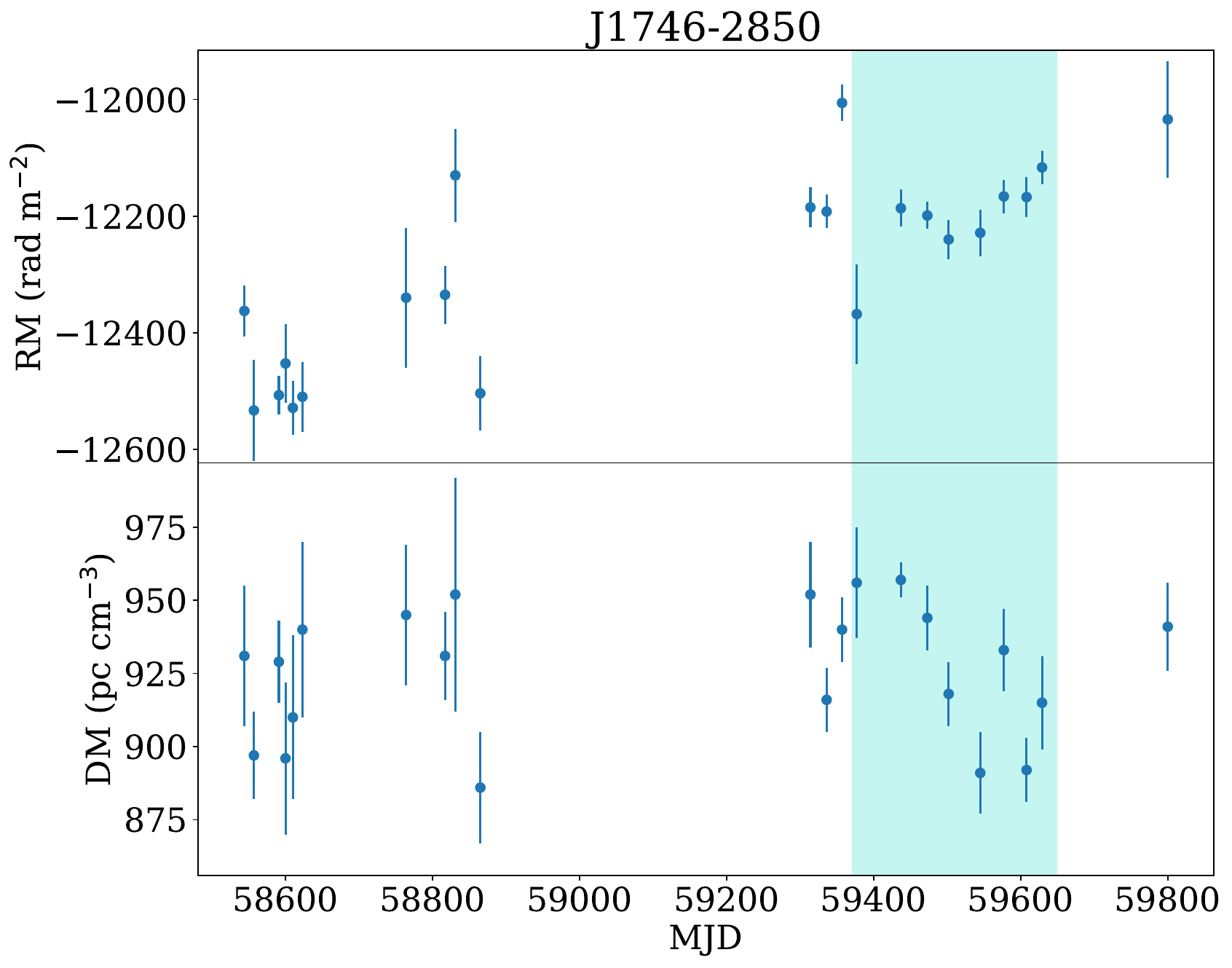}
	\,
	\includegraphics[width=0.48\textwidth]{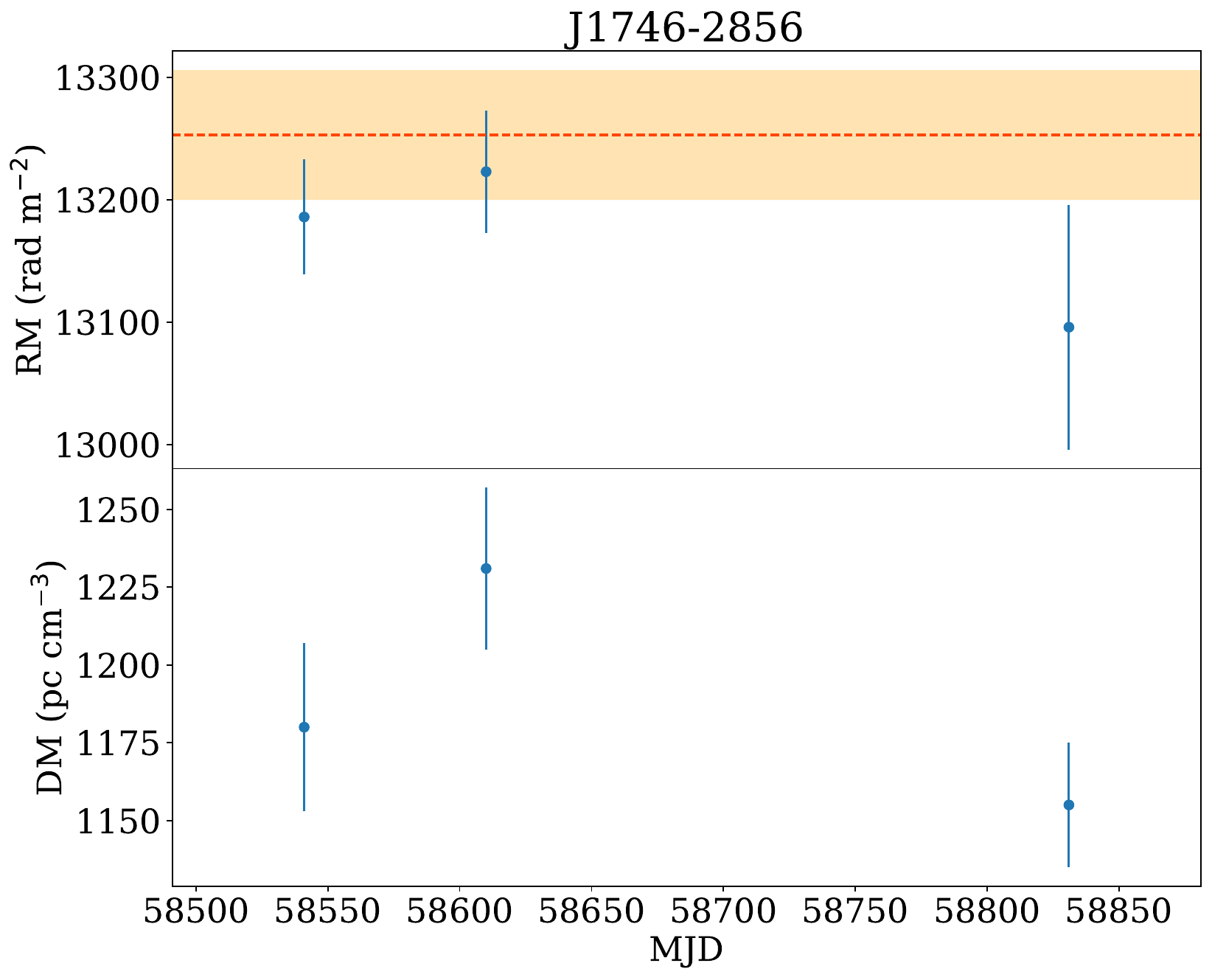}	
	\,
	\includegraphics[width=0.48\textwidth]{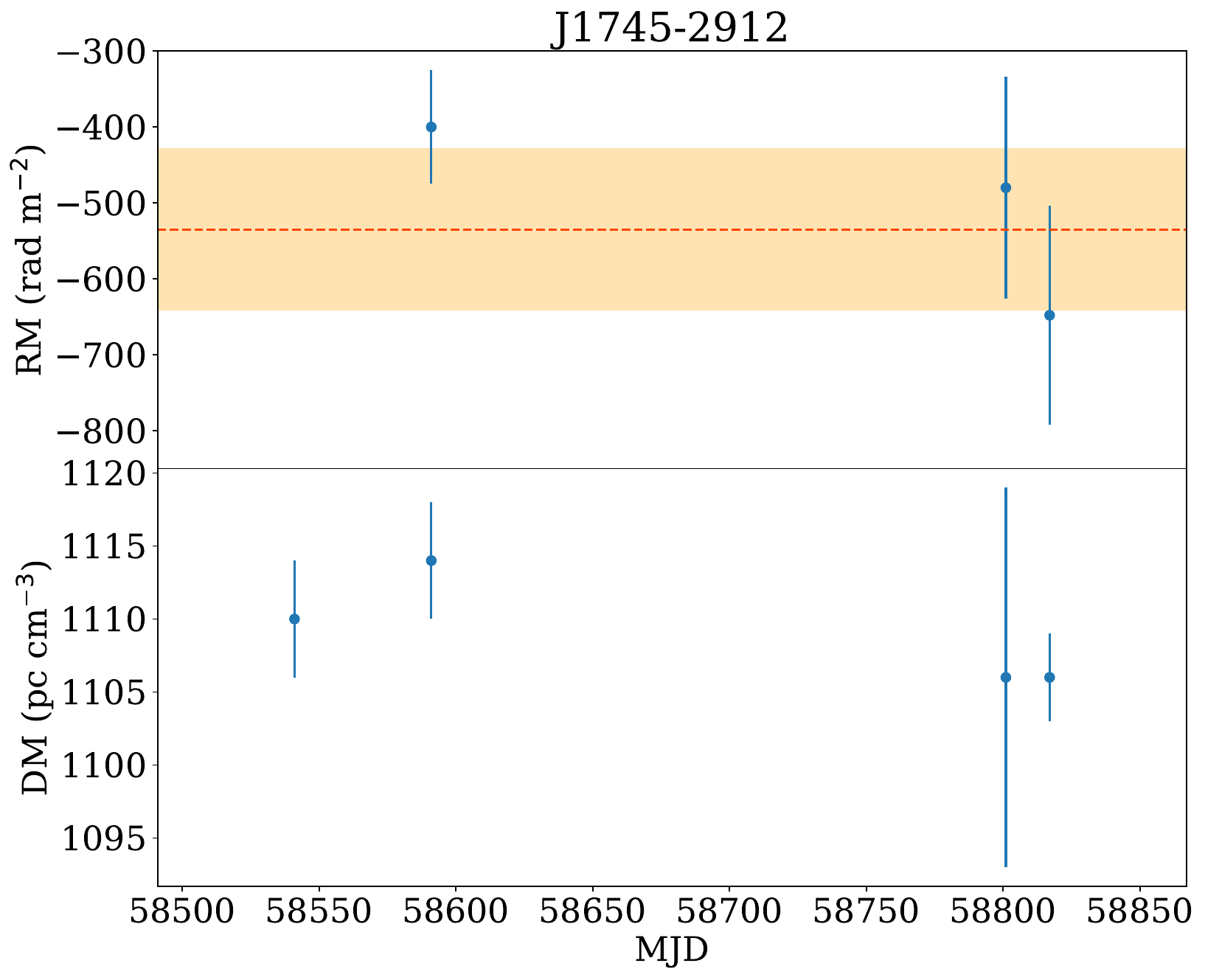}
  	\caption{Variations of RM and DM for the observed pulsars as a function of the MJD of the observation. The errorbars show the 1$\sigma$ uncertainty interval. The red dashed line and the orange region show the value and 1$\sigma$ uncertainty interval of the RM measured in {\protect \cite{Schnitzeler2016}}. The region highlighted in cyan in the plot for J1746$-$2850 shows the event described in the text where the RM increases by $\sim$ 200 rad m$^{-2}$ while the DM decreases by $\sim$ 50 pc cm$^{-3}$.
   }
  	\label{fig:RM_DM_variations}
\end{figure*}

The RM-corrected polarization profiles of PSR J1746$-$2849, PSR J1746$-$2850, PSR J1746$-$2856 and PSR J1745$-$2912 are shown in Fig. \ref{fig:profiles}.
The values of the DMs and RMs at each observation are shown in Fig. \ref{fig:RM_DM_variations} and in Table \ref{tab:results}.

For comparison purposes we show also the values of RM reported in \cite{Schnitzeler2016} for PSR J1746$-$2849, PSR J1746$-$2856 and PSR J1745$-$2912. Those values reported in Table \ref{tab:results} and shown as a dashed red line with an orange 1$\sigma$ region in Fig. \ref{fig:RM_DM_variations} are compatible with the variability reported in this manuscript. This implies that, over 7 years, the average RM values for these pulsars have remained roughly constant and that the variability occurs on shorter timescales.

For PSR J1746$-$2850 this is the first published measurement of RM. The  median value of RM is $\sim -12234$ \radm with an error estimated using the median of the absolute deviation from the median of 181 \radm. This makes it the pulsar with third highest absolute value of RM after PSR J1745$-$2900 and PSR J1746$-$2856. Because of its implied high magnetic field, flat spectrum and transient behaviour, PSR J1746$-$2850 has been compared to radio loud magnetars \citep{Dexter2017}. Thanks to the polarization observations we can calculate the linear polarization percentage and compare it with the
very high ($\sim 80-100$ percent) linear polarization of magnetars \citep{Camilo2006,Camilo2008,eatough2013,Levin2012}. The linear polarization of this pulsar is found to be 43 percent. This challenges the classification of PSR J1746$-$2850 as a magnetar but reinforces the idea that it might be a transitional object between a rotation-powered pulsar and a magnetar.

As seen in Fig. \ref{fig:pulsars_positions}, PSR J1746$-$2850 is located close to the Arc NTF in a region dominated by the Sickle HII region G0.18-0.04 \citep{Yusef-zadeh1984}. The young age of the pulsar derived from the timing properties ($T\simeq 13$ kyr, \citealt{Deneva2009}) suggests that it might have originated from the nearby Quintuplet or Arches clusters \citep{Dexter2017} and might be still located in the vicinity of the gas rich HII regions visible in Fig. \ref{fig:pulsars_positions}. The high absolute value of RM is compatible with this position given the very high electron densities of these clouds, $\sim 300-400$ \pccm \citep{Simpson2007}, and the large expected magnetic field in the region, between $100$ $\mu$G and 1 mG \citep{Ferriere2009,Oka2019}.   

The behaviours of DM and RM are expected to be different for sources close to the Galactic centre. While the RM is primarily affected by local screens in regions with high magnetic fields, most of the DM arises from the collective effect of the electron along the entire line of sight \citep{eatough2013,Desvignes2018}. 
We test if there are significant variations in the RMs and DMs of the pulsars or if they are compatible with a single value. We perform a $\chi^2$ test with the null-hypothesis that the RMs and DMs of the pulsars are compatible with a single value using a threshold p-value of 0.001. For J1746-2849 we obtain a p-value of 0.7 for the RMs and 0.5 for the DMs, for J1746-2850 we find $<10^{-5}$ for the RMs and $<10^{-5}$ for the DMs, for J1746-2856 we find 0.5 for the RMs and 0.07 for the DMs, while for J1745-2912 we find 0.3 for the RMs and 0.4 for the DMs. Therefore, only pulsar J1746-2850 shows significant variations in RM and DM.
The variations in RM can occur even on timescales of $\sim 10$ days, the smallest timescales at which we observe. This suggests that the variability could occur even at smaller timescales. The variations in RM are smaller than what has been observed for PSR J1745$-$2900 over similar timescales \citep{Desvignes2018}. This implies that the magnetic field is significantly stronger in the vicinity of Sgr A$^{*}$ where PSR J1745$-$2900 is located compared to the location of the pulsars observed in this work.  

An interesting event, highlighted in light blue in Fig. \ref{fig:RM_DM_variations} occurred around MJD 59400 and lasting $\sim 100$ days for PSR J1746$-$2850 where the RM was measured to be  monotonically increasing by $\sim 200$ \radm while the DM decreased by $\sim 50$ \pccm. 
Similar secular variations of RM over a period of a few months have already been observed in the Galactic centre magnetar \citep{Desvignes2018} and in FRB 121102 \citep{Hilmarsson2021} and FRB 20180916B \citep{Mckinven2022}. However, in the case of PSR J1746$-$2850, we also observe a simultaneous decrease, on average, of the DM. The large and simultaneous RM and DM variations suggest that the event arises from a local source possibly located within the CMZ. Under the assumption that the contribution of the gas along the entire line of sight to the pulsar remains constant, we can estimate the average component of the magnetic field parallel to the line of sight within the local source with the equation:



\begin{equation}
    B_{\parallel} \sim 1.23 \frac{\rm RM_{\rm end} -RM_{\rm start}}{\rm DM_{\rm end}-DM_{\rm start}} \sim - 5 \, \mu G,
\end{equation}
where RM$_{\rm start}$ and  DM$_{\rm start}$ are the RM and DM at the beginning of the event and RM$_{\rm end}$ and  DM$_{\rm end}$ are the RM and DM at the end.

This value of magnetic field is significantly lower than the value of $0.1-1$ mG expected for the CMZ \citep{Ferriere2009,Oka2019}. The low value of the projected magnetic field together with the strong variability of electron density, as suggested by the DM, suggests that either the observed variability is happening in a gas-rich region with low magnetic field or that the magnetic field is mostly perpendicular to the line of sight.

\subsection{RM and DM structure functions for PSR J1746$-$2850}

In the case of PSR J1746$-$2850 we can check if the variations are caused by a turbulent medium by looking at the second-order structure function \citep[SF,][]{Lazarian2016}. The RM SF is defined as:

\begin{equation}
    D_{\rm RM} (\tau)= \langle [{\rm RM}(t+\tau)-{\rm RM}(t)]^2\rangle,
\end{equation}
where $t$ is the time of each observation, $\tau$ is the lag between two different observations and the angle brackets mean the average between all pairs with the same temporal difference.

The time difference between the observations can be considered as a proxy of the angular separation between the position of the pulsar at different observations.
The pulsar could have proper motion of several hundreds of km s$^{-1}$ \citep{Lyne1994,Verbunt2017} while
the velocity of the gas in the Sickle HII region is measured to be $\sim 30-80$ km s$^{-1}$ \citep{Butterfield2018}. This means that the line of sight to the pulsar will traverse different parts of gas with different density and magnetic fields at different times. 
In the case of a turbulent medium, we expect the SF to follow either a single power-law or a broken power-law \citep{Minter1996,Lazarian2016}.

\begin{figure}
\centering
	\includegraphics[width=0.4\textwidth]{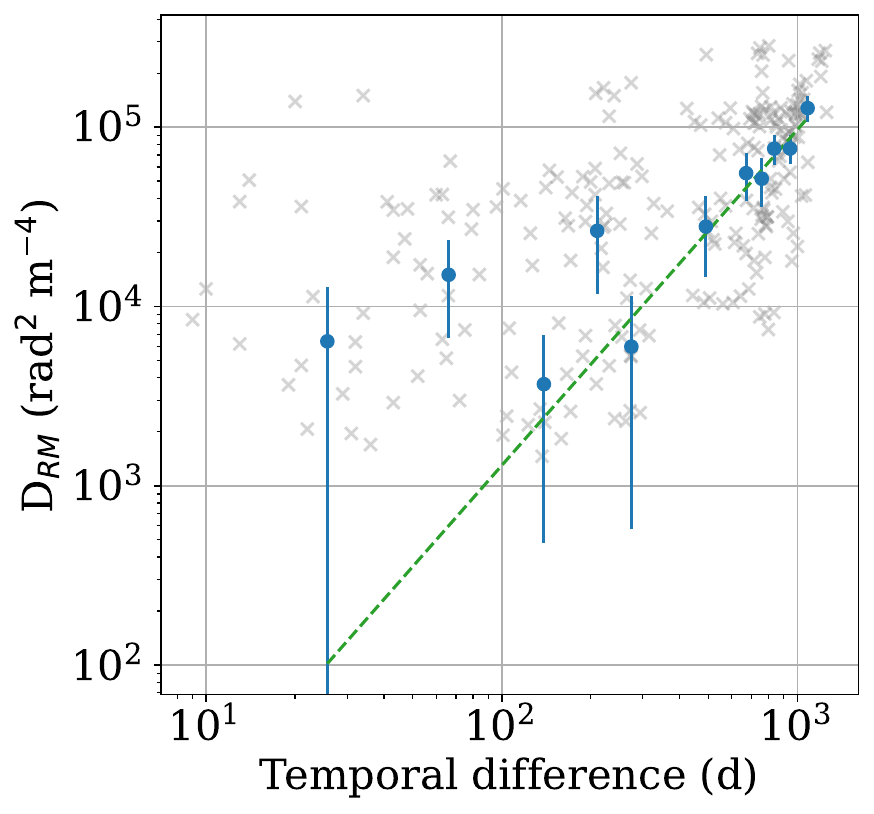}
	\\
	\includegraphics[width=0.4\textwidth]{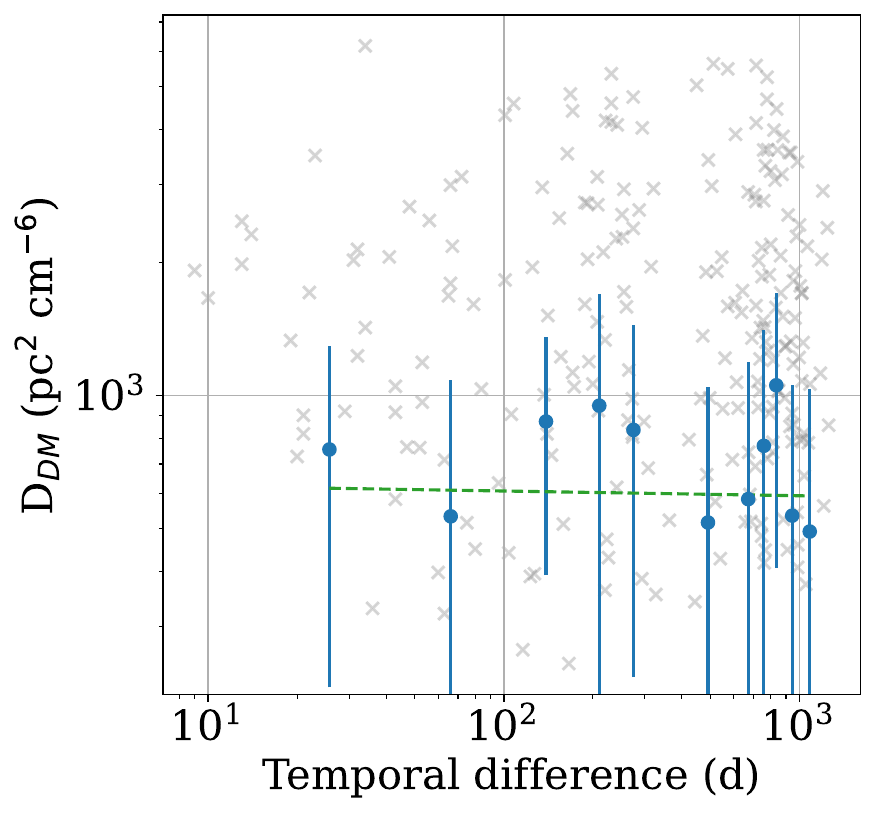}
	\,
  	\caption{ SFs of RM (above) and DM  (below) for PSR J1746$-$2850. The gray crosses are the values for each pair while filled blue circles represent the averaged values over 21 pairs after removing the noise level. 
   The dashed green lines show the best-fitting power-laws. 
   The best-fitting power-law index for the RM SF is $1.87_{-0.3}^{+0.4}$ while for the DM it is $-0.01 \pm 0.34$. The error bars show the 68 percent confidence interval.}
  	\label{fig:RM_DM_SFs}
\end{figure}

To measure the RM SF we first estimate the square difference of all RM pairs from different observations. We then sort the pairs as a function of time difference between the observations and group the 231 pairs together in 11 bins containing each 21 pairs. 
The value and the error of the SF for each bin is measured by taking the median and the standard error of the median with a Monte Carlo method.
We report the RM SF in top panel of Fig. \ref{fig:RM_DM_SFs}.

We further verify the statistical significance of the SF by simulating the effects of the white noise. We repeat the same Monte Carlo extraction as described before but we use the same value of RM for each pulsar. The resulting SF represents the noise level given by the errors on the RM and is subtracted from the true SF before doing the fit.

We perform a fit to the data with a single power-law. 
The fits are based on the Markov Chain Monte Carlo code \texttt{emcee} \citep{Foreman-Mackey2013}. The free parameters are the power-law index and the value of the SF at a separation of 1 day. For the power-law indices we used a flat prior between $-2$ and 4. 
The 68 percent confidence interval of the power-law index is $1.87_{-0.3}^{+0.4}$ and for the value at 1 day it is $0.23_{-0.01}^{+1.44}$ rad$^{2}$ m$^{-4}$.
The best-fitting power-law is shown in the top panel of Fig. \ref{fig:RM_DM_SFs} as the dashed green line. In the bins corresponding to the smallest temporal differences, the best fitting model goes below the noise limit meaning that these points are dominated by the noise.
The power-law index is compatible to what is expected by isotropic Kolmogorov turbulence (that predicts a value of 5/3, e.g. \citealt{Lazarian2016}). The strong variability not seen in other pulsar along the Galactic disk suggests that the turbulence is located in the CMZ close to the pulsar.
The power-law index is larger but still compatible given the large errors with the value of $1.23 \pm 0.13$ measured for PSR J1745$-$2900 \citep{Desvignes2018}.

We repeat the same process for the DM and show the results in the bottom panel of Fig. \ref{fig:RM_DM_SFs}. If we attempt to perform a fit to the observed values we get a 68 percent confidence interval for the power-law index of $-0.01\pm 0.34$ and for the value at 1 day it is $540_{-500}^{+1600}$ pc$^{2}$ cm$^{-6}$. We notice that the observed values of the SF are close to the sensitivity limit for a large number of bins. This suggests the result is highly impacted by the errors on the measurements. The real underlying SF could be smaller and the apparent flat behaviour could be a consequence of the large errors. However, a similar flat behaviour has also been observed in PSR J1745$-$2900 \citep{Desvignes2018}.


\subsection{Implications for magnetic fields in the Radio Arc}

\begin{figure*}
\centering
	\includegraphics[width=0.8\textwidth]{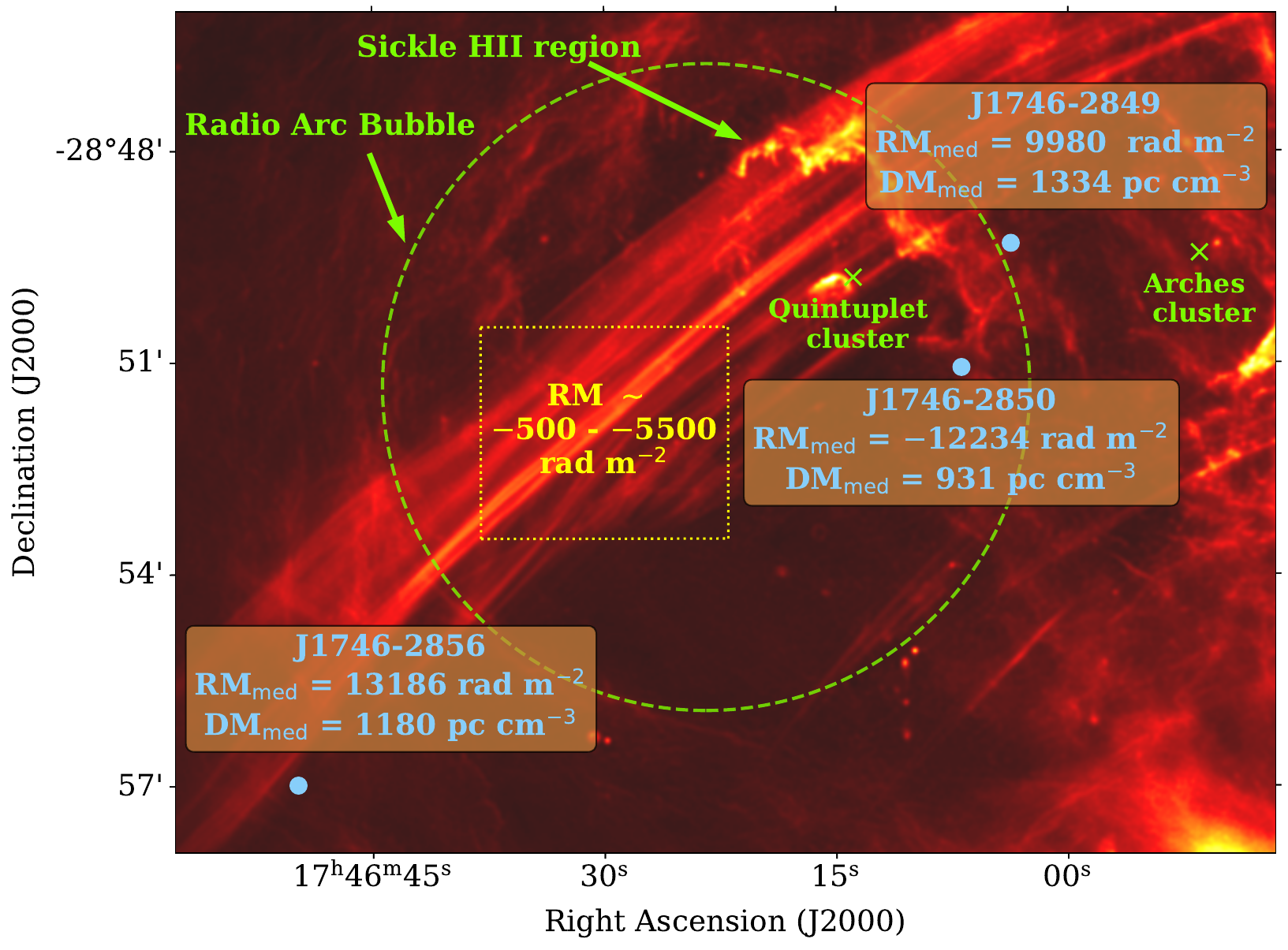}
    \caption{Enlargement of Fig. \ref{fig:pulsars_positions} in the region of the Radio Arc. The background comes from L-band observations obtained at MeerKAT \citep{Heywood2022}. The position of the Sickle HII region and the Radio Arc Bubble are shown in green together with the position of the Quintuplet and Arches clusters. The yellow dotted square corresponds to the region of the Radio Arc where polarization is strong enough to allow a measurement of RM {\protect \citep{Pare2019, Pare2021}}. The range of RMs measured in those works are reported inside the square in yellow. At the distance of Sgr A$^*$, the physical size of the image is 32x26 pc. 
    }
  	\label{fig:pulsars_positions_zoom}
\end{figure*}

Figure \ref{fig:pulsars_positions_zoom} shows the position of the pulsars located within the Radio Arc with respect to the main radio continuum features. We show the position of the Sickle HII region G0.18-0.04 \citep{Yusef-zadeh1984} and the Radio Arc Bubble \citep{Simpson2007, Pare2021} that is suggested to be driven by the outflow of the nearby Quintuplet cluster.

PSR J1746$-$2849 and PSR J1746$-$2850 are located close to each other with an angular separation of 1.9 arcmin, corresponding to a projected physical separation of 4.5 pc at the distance of Galactic Centre. Both pulsars are located close to the Sickle HII region near the border of the Radio Arc Bubble. Despite the small angular difference, the DM difference between the two pulsars is $\sim 400$ \pccm while the RM difference is $\sim 22000$ \radm. The higher value of DM of PSR J1746$-$2849 indicates either a position in a denser environment or that it is located at a larger distance from us. The electron density in the Sickle was determined to be $\sim 220$ cm$^{-3}$ from radio continuum observations \citep{Lang1997} and $\sim 300-400$ cm$^{-3}$ from spectral lines observations \citep{Simpson2007}. The electron density is related to the DM by the equation \citep{Handbook}:
\begin{equation}
    {\rm DM}= \int_0^d n_e \,{\rm d}l\, {\rm [pc \, cm^{-3}]},
\end{equation}
where $d$ is the distance from the observer to the pulsar expressed in pc and $n_e$ is the electron density expressed in cm$^{-3}$. This implies that, within the densest regions, a difference in DM of $\sim 400$ \pccm can occur over only 1 or 2 pc.

Given the close proximity in projection of these pulsars it is possible that there is a common Faraday screen in front of them. Unfortunately, we do not detect significant variability in RM for J1746$-$2849. Even if RM variations of the same magnitude as for J1746$-$2850 are present, they would not be detectable due to the large uncertainties. 

As an order of magnitude estimate, we can probe the component of the magnetic field parallel to the line of sight by comparing the DM and RM of these two pulsars. In the assumption that the foreground contribution to DM and RM is the same, the differences would be caused entirely by the extra ionized gas between the position of PSR J1746$-$2850 and PSR J1746$-$2849. The parallel component of the magnetic field in this region needed to generate this difference in RM is:

\begin{equation}
    B_{\parallel} \sim 1.23 \frac{\rm RM_{J1746-2849}-RM_{J1746-2850}}{\rm DM_{J1746-2849}-DM_{J1746-2850}} \sim 70 \, \mu G.
\end{equation}

Given that this is only the average value of the parallel component of the magnetic field averaged over the line of sight, the magnetic field will likely be stronger. This would support the idea that the magnetic field within the Arc NTF is of the order of 0.1 - 1 mG \citep{Ferriere2009,Oka2019}. 

We can compare the values of RM of the pulsars and previous studies of the Radio Arc. Polarization observations of the entire region have revealed that the polarization is concentrated in a region marked by the dotted box shown in Fig. \ref{fig:pulsars_positions_zoom}. In this box the RMs vary from $\sim -500$ to $\sim -5500$ \radm \citep{Pare2019,Pare2021}. The lack of detectable polarization in the region close to the pulsars could be explained by variations of magnetic field strength and orientation within the observing telescope beam, different polarization angles averaged over the beam \citep{Burn1966,Pare2019} or by thermal emission from large electron densities in HII regions \citep{Pare2021}. These depolarization effects do not affect the pulsar observations as much. The reason is that, when observing a pulsar, we fold the time series of the observation at the specific rotational period of the pulsar. By doing so we add coherently only the signal from the pulsar while any other source in the field of view is added incoherently and contributes to a uniform background. By subtracting the background we are able to isolate the signal and polarization of the pulsar and remove the other sources that could lead to depolarization. This effect could be exploited when looking for new pulsars in such dense environments as they are likely to be the only polarized sources in the field of view.

The values of the RMs observed in pulsars are significantly larger than the one observed in the imaging observations of the Radio Arc. In particular, the very large RM difference between PSR J1746$-$2849 and PSR J1746$-$2850 and the RM variability of more than 300 \radm over a few months suggest that the magnetic environment surrounding the pulsars is more complicated and variable than the region where polarization is visible in radio imaging.

\subsection{Comparison with FRBs}

RM variations as large as the ones in our sample are rare within the known pulsars with a few exceptions like PSR J1745$-$2900 that shows variations of $\sim 3000$ \radm \citep{Desvignes2018} and some binary systems around gas-shedding stars \citep{Johnston2005,Li2022}. The only other group of objects that show similar RM variations are FRBs, e.g. FRB 20201124A with variations of $\sim 500$ \radm  \citep{Hilmarsson2021b,Wang2022}, FRB 121102 with variations of $\sim 30000$ \radm  \citep{Hilmarsson2021} FRB 20190520B with variations of $\sim 40000$ \radm  \citep{Anna-Thomas2023}. 
 One interesting case  is FRB 20180916B \citep{Mckinven2022} that shows a secular increase in RM of $\sim 50$ \radm over a period of 8 months. We see a similar secular variation for PSR J1746$-$2850 in the event occurring around MJD 59400. 
 However, the RM SF of FRB 20180916B has a flat  power-law index of $\sim 0.3$ \citep{Mckinven2022} compared to the steep value of $1.87_{-0.3}^{+0.4}$ for PSR J1746$-$2850. Furthermore, FRB 20180916B is located in a region of the host galaxy with low star formation rate and low H$\alpha$ luminosity \citep{Tendulkar2021} which is quite different from the bright and active CMZ.
 

\section{Conclusions}

We observed four of the six pulsars closest to the Galactic centre, PSR J1746$-$2849, PSR J1746$-$2850, PSR J1746$-$2856 and PSR J1745$-$2912 with the Effelsberg 100-m radio telescope in order to study the DM and RM variations over time. This complements high cadence observations of PSR J1745$-$2900 that were presented in \cite{Desvignes2018}. We report for the first time the value of RM of PSR J1746$-$2850 of $-12234 \pm 181$ \radm, the third highest absolute value of RM of any known pulsar. Over the time of the observations, this pulsar shows large variations of RM of around $300-400$ \radm. These are among the strongest variations in RM known for pulsars but are still smaller than those of PSR J1745$-$2900 \citep{Desvignes2018} and show similarities with the variations observed in some FRBs \citep{Mckinven2022, Hilmarsson2021}.
The DM variations are smaller and in most cases compatible with a constant value except for PSR J1746$-$2850. For this pulsar, we observe an event occurring around MJD 59400 where the DM decreases systematically by about 50 \pccm while the RM increases simultaneously by about 200 \radm. While DM variations of this magnitude are rare in pulsars, they have been observed in FRB 20190529B \citep{Anna-Thomas2023} and, to a lower extent also in FRB 121102 \citep{Wang2022B}.

PSR J1746$-$2850 is the only pulsar for which we have enough observations to analyse the DM and RM SFs. 
The RM SF shows a growing trend with longer time separations with a power-law index of $1.87_{-0.3}^{+0.4}$, a value that is compatible with the expected value of 5/3 in case of three dimensional isotropic turbulence. If real, this turbulence is likely to be related with the Sickle HII region or with the Radio Arc bubble. The DM SF, on the other hand, shows a flat behaviour but the errors are too large for this to be conclusive.

The very large difference of RM between PSR J1746$-$2849 and PSR J1746$-$2850 of $\sim 22000$ \radm despite being located only 1.9 arcmin apart suggests the presence of a very large magnetic field in the region that could be stronger than $\sim 70$ $\mu$G. This, combined with the large RM variations of over $\sim 300$ \radm, could give an explanation to the depolarization in the imaging observations of the Sickle HII region.

Future observations of the pulsars close to the Galactic centre would allow us to determine the RM and DM SFs and to study in more detail the behaviour of PSR J1746$-$2850. Observations with the MeerKAT telescope at the proposed S-band (1.75-3.55 GHz) would allow a better determination of the DM and RM variations thanks to the higher elevation in the sky and the higher sensitivity. With the advent of this new facility we expect a potential discovery of more pulsars in the region that could provide precious information of the magneto-ionic properties close to the Galactic centre even in areas where continuum observations do not show polarization.

\section*{Acknowledgements}
This work was based on observations with the $100\,{\rm m}$ telescope of the Max-Planck-Institut f\"{u}r Radioastronomie at Effelsberg. RPE Funded by Chinese Academy of Sciences President’s International Fellowship Initiative. Grant No. 2021FSM0004. FA, AN, GD, RW, PT, MK, RPE, RK, KL, and LS acknowledge the financial support by the European Research Council for the ERC Synergy grant Black Hole Cam under contract no. 610058. This work is supported by the Max-Planck Society as part of the "LEGACY" collaboration on low-frequency gravitational wave astronomy. 
R.S.W. was supported by an appointment to the NASA Postdoctoral Program at the Jet Propulsion Laboratory, administered by Oak Ridge Associated Universities under contract with NASA.
Part of this research was carried out at the Jet Propulsion Laboratory, California Institute of Technology, under a contract with the National Aeronautics and Space Administration.
\section*{Data Availability}

The data underlying this article will be shared upon reasonable request to the corresponding author.



\bibliographystyle{mnras}
\bibliography{biblio} 

\begin{thebibliography}{}
\makeatletter
\relax
\def\mn@urlcharsother{\let\do\@makeother \do\$\do\&\do\#\do\^\do\_\do\%\do\~}
\def\mn@doi{\begingroup\mn@urlcharsother \@ifnextchar [ {\mn@doi@}
  {\mn@doi@[]}}
\def\mn@doi@[#1]#2{\def\@tempa{#1}\ifx\@tempa\@empty \href
  {http://dx.doi.org/#2} {doi:#2}\else \href {http://dx.doi.org/#2} {#1}\fi
  \endgroup}
\def\mn@eprint#1#2{\mn@eprint@#1:#2::\@nil}
\def\mn@eprint@arXiv#1{\href {http://arxiv.org/abs/#1} {{\tt arXiv:#1}}}
\def\mn@eprint@dblp#1{\href {http://dblp.uni-trier.de/rec/bibtex/#1.xml}
  {dblp:#1}}
\def\mn@eprint@#1:#2:#3:#4\@nil{\def\@tempa {#1}\def\@tempb {#2}\def\@tempc
  {#3}\ifx \@tempc \@empty \let \@tempc \@tempb \let \@tempb \@tempa \fi \ifx
  \@tempb \@empty \def\@tempb {arXiv}\fi \@ifundefined
  {mn@eprint@\@tempb}{\@tempb:\@tempc}{\expandafter \expandafter \csname
  mn@eprint@\@tempb\endcsname \expandafter{\@tempc}}}

\bibitem[\protect\citeauthoryear{{Abbate}, {Possenti}, {Tiburzi}, {Barr}, {van
  Straten}, {Ridolfi}  \& {Freire}}{{Abbate} et~al.}{2020}]{Abbate2020}
{Abbate} F.,  {Possenti} A.,  {Tiburzi} C.,  {Barr} E.,  {van Straten} W.,
  {Ridolfi} A.,   {Freire} P.,  2020, \mn@doi [Nature Astronomy]
  {10.1038/s41550-020-1030-6}, \href
  {https://ui.adsabs.harvard.edu/abs/2020NatAs...4..704A} {4, 704}

\bibitem[\protect\citeauthoryear{Anna-Thomas et~al.,}{Anna-Thomas
  et~al.}{2023}]{Anna-Thomas2023}
Anna-Thomas R.,  et~al., 2023, \mn@doi [Science] {10.1126/science.abo6526},
  380, 599

\bibitem[\protect\citeauthoryear{{Brentjens} \& {de Bruyn}}{{Brentjens} \& {de
  Bruyn}}{2005}]{Brentjens2005}
{Brentjens} M.~A.,  {de Bruyn} A.~G.,  2005, \mn@doi [\aap]
  {10.1051/0004-6361:20052990}, \href
  {https://ui.adsabs.harvard.edu/abs/2005A&A...441.1217B} {441, 1217}

\bibitem[\protect\citeauthoryear{{Burn}}{{Burn}}{1966}]{Burn1966}
{Burn} B.~J.,  1966, \mn@doi [\mnras] {10.1093/mnras/133.1.67}, \href
  {https://ui.adsabs.harvard.edu/abs/1966MNRAS.133...67B} {133, 67}

\bibitem[\protect\citeauthoryear{{Butterfield}, {Lang}, {Morris}, {Mills}  \&
  {Ott}}{{Butterfield} et~al.}{2018}]{Butterfield2018}
{Butterfield} N.,  {Lang} C.~C.,  {Morris} M.,  {Mills} E. A.~C.,   {Ott} J.,
  2018, \mn@doi [\apj] {10.3847/1538-4357/aa886e}, \href
  {https://ui.adsabs.harvard.edu/abs/2018ApJ...852...11B} {852, 11}

\bibitem[\protect\citeauthoryear{{Camilo}, {Ransom}, {Halpern}, {Reynolds},
  {Helfand}, {Zimmerman}  \& {Sarkissian}}{{Camilo} et~al.}{2006}]{Camilo2006}
{Camilo} F.,  {Ransom} S.~M.,  {Halpern} J.~P.,  {Reynolds} J.,  {Helfand}
  D.~J.,  {Zimmerman} N.,   {Sarkissian} J.,  2006, \mn@doi [\nat]
  {10.1038/nature04986}, \href
  {https://ui.adsabs.harvard.edu/abs/2006Natur.442..892C} {442, 892}

\bibitem[\protect\citeauthoryear{{Camilo}, {Reynolds}, {Johnston}, {Halpern}
  \& {Ransom}}{{Camilo} et~al.}{2008}]{Camilo2008}
{Camilo} F.,  {Reynolds} J.,  {Johnston} S.,  {Halpern} J.~P.,   {Ransom}
  S.~M.,  2008, \mn@doi [\apj] {10.1086/587054}, \href
  {https://ui.adsabs.harvard.edu/abs/2008ApJ...679..681C} {679, 681}

\bibitem[\protect\citeauthoryear{{Cordes} \& {Chatterjee}}{{Cordes} \&
  {Chatterjee}}{2019}]{Cordes2019}
{Cordes} J.~M.,  {Chatterjee} S.,  2019, \mn@doi [\araa]
  {10.1146/annurev-astro-091918-104501}, \href
  {https://ui.adsabs.harvard.edu/abs/2019ARA&A..57..417C} {57, 417}

\bibitem[\protect\citeauthoryear{{Deneva}, {Cordes}  \& {Lazio}}{{Deneva}
  et~al.}{2009}]{Deneva2009}
{Deneva} J.~S.,  {Cordes} J.~M.,   {Lazio} T.~J.~W.,  2009, \mn@doi [\apjl]
  {10.1088/0004-637X/702/2/L177}, \href
  {https://ui.adsabs.harvard.edu/abs/2009ApJ...702L.177D} {702, L177}

\bibitem[\protect\citeauthoryear{{Desvignes} et~al.,}{{Desvignes}
  et~al.}{2018}]{Desvignes2018}
{Desvignes} G.,  et~al., 2018, \mn@doi [\apjl] {10.3847/2041-8213/aaa2f8},
  \href {https://ui.adsabs.harvard.edu/abs/2018ApJ...852L..12D} {852, L12}

\bibitem[\protect\citeauthoryear{{Dexter} et~al.,}{{Dexter}
  et~al.}{2017}]{Dexter2017}
{Dexter} J.,  et~al., 2017, \mn@doi [\mnras] {10.1093/mnras/stx583}, \href
  {https://ui.adsabs.harvard.edu/abs/2017MNRAS.468.1486D} {468, 1486}

\bibitem[\protect\citeauthoryear{{Eatough} et~al.,}{{Eatough}
  et~al.}{2013}]{eatough2013}
{Eatough} R.~P.,  et~al., 2013, \mn@doi [\nat] {10.1038/nature12499}, \href
  {https://ui.adsabs.harvard.edu/abs/2013Natur.501..391E} {501, 391}

\bibitem[\protect\citeauthoryear{{Ferri{\`e}re}}{{Ferri{\`e}re}}{2009}]{Ferriere2009}
{Ferri{\`e}re} K.,  2009, \mn@doi [\aap] {10.1051/0004-6361/200912617}, \href
  {https://ui.adsabs.harvard.edu/abs/2009A&A...505.1183F} {505, 1183}

\bibitem[\protect\citeauthoryear{{Ferri{\`e}re}, {Gillard}  \&
  {Jean}}{{Ferri{\`e}re} et~al.}{2007}]{Ferriere2007}
{Ferri{\`e}re} K.,  {Gillard} W.,   {Jean} P.,  2007, \mn@doi [\aap]
  {10.1051/0004-6361:20066992}, \href
  {https://ui.adsabs.harvard.edu/abs/2007A&A...467..611F} {467, 611}

\bibitem[\protect\citeauthoryear{{Foreman-Mackey}, {Hogg}, {Lang}  \&
  {Goodman}}{{Foreman-Mackey} et~al.}{2013}]{Foreman-Mackey2013}
{Foreman-Mackey} D.,  {Hogg} D.~W.,  {Lang} D.,   {Goodman} J.,  2013, \mn@doi
  [Publications of the Astronomical Society of the Pacific] {10.1086/670067},
  \href {https://ui.adsabs.harvard.edu/abs/2013PASP..125..306F} {125, 306}

\bibitem[\protect\citeauthoryear{{Hales}, {Gaensler}, {Norris}  \&
  {Middelberg}}{{Hales} et~al.}{2012}]{Hales2012}
{Hales} C.~A.,  {Gaensler} B.~M.,  {Norris} R.~P.,   {Middelberg} E.,  2012,
  \mn@doi [\mnras] {10.1111/j.1365-2966.2012.21372.x}, \href
  {https://ui.adsabs.harvard.edu/abs/2012MNRAS.424.2160H} {424, 2160}

\bibitem[\protect\citeauthoryear{{Han}, {Manchester}, {Lyne}, {Qiao}  \& {van
  Straten}}{{Han} et~al.}{2006}]{Han2006}
{Han} J.~L.,  {Manchester} R.~N.,  {Lyne} A.~G.,  {Qiao} G.~J.,   {van Straten}
  W.,  2006, \mn@doi [\apj] {10.1086/501444}, \href
  {https://ui.adsabs.harvard.edu/abs/2006ApJ...642..868H} {642, 868}

\bibitem[\protect\citeauthoryear{{Han}, {Manchester}, {van Straten}  \&
  {Demorest}}{{Han} et~al.}{2018}]{Han2018}
{Han} J.~L.,  {Manchester} R.~N.,  {van Straten} W.,   {Demorest} P.,  2018,
  \mn@doi [\apjs] {10.3847/1538-4365/aa9c45}, \href
  {https://ui.adsabs.harvard.edu/abs/2018ApJS..234...11H} {234, 11}

\bibitem[\protect\citeauthoryear{{Heywood} et~al.,}{{Heywood}
  et~al.}{2022}]{Heywood2022}
{Heywood} I.,  et~al., 2022, \mn@doi [\apj] {10.3847/1538-4357/ac449a}, \href
  {https://ui.adsabs.harvard.edu/abs/2022ApJ...925..165H} {925, 165}

\bibitem[\protect\citeauthoryear{{Hilmarsson}, {Spitler}, {Main}  \&
  {Li}}{{Hilmarsson} et~al.}{2021a}]{Hilmarsson2021b}
{Hilmarsson} G.~H.,  {Spitler} L.~G.,  {Main} R.~A.,   {Li} D.~Z.,  2021a,
  \mn@doi [\mnras] {10.1093/mnras/stab2936}, \href
  {https://ui.adsabs.harvard.edu/abs/2021MNRAS.508.5354H} {508, 5354}

\bibitem[\protect\citeauthoryear{{Hilmarsson} et~al.,}{{Hilmarsson}
  et~al.}{2021b}]{Hilmarsson2021}
{Hilmarsson} G.~H.,  et~al., 2021b, \mn@doi [\apjl] {10.3847/2041-8213/abdec0},
  \href {https://ui.adsabs.harvard.edu/abs/2021ApJ...908L..10H} {908, L10}

\bibitem[\protect\citeauthoryear{{Hobbs}, {Edwards}  \& {Manchester}}{{Hobbs}
  et~al.}{2006}]{Hobbs2006}
{Hobbs} G.~B.,  {Edwards} R.~T.,   {Manchester} R.~N.,  2006, \mn@doi [MNRAS]
  {10.1111/j.1365-2966.2006.10302.x}, \href
  {https://ui.adsabs.harvard.edu/abs/2006MNRAS.369..655H} {369, 655}

\bibitem[\protect\citeauthoryear{{Hotan}, {van Straten}  \&
  {Manchester}}{{Hotan} et~al.}{2004}]{Hotan2004}
{Hotan} A.~W.,  {van Straten} W.,   {Manchester} R.~N.,  2004, \mn@doi
  [Publications of the Astronomical Society of Australia] {10.1071/AS04022},
  \href {https://ui.adsabs.harvard.edu/abs/2004PASA...21..302H} {21, 302}

\bibitem[\protect\citeauthoryear{{Johnston}, {Ball}, {Wang}  \&
  {Manchester}}{{Johnston} et~al.}{2005}]{Johnston2005}
{Johnston} S.,  {Ball} L.,  {Wang} N.,   {Manchester} R.~N.,  2005, \mn@doi
  [\mnras] {10.1111/j.1365-2966.2005.08854.x}, \href
  {https://ui.adsabs.harvard.edu/abs/2005MNRAS.358.1069J} {358, 1069}

\bibitem[\protect\citeauthoryear{{Johnston}, {Kramer}, {Lorimer}, {Lyne},
  {McLaughlin}, {Klein}  \& {Manchester}}{{Johnston}
  et~al.}{2006}]{Johnston2006}
{Johnston} S.,  {Kramer} M.,  {Lorimer} D.~R.,  {Lyne} A.~G.,  {McLaughlin} M.,
   {Klein} B.,   {Manchester} R.~N.,  2006, \mn@doi [\mnras]
  {10.1111/j.1745-3933.2006.00232.x}, \href
  {https://ui.adsabs.harvard.edu/abs/2006MNRAS.373L...6J} {373, L6}

\bibitem[\protect\citeauthoryear{{Johnston} et~al.,}{{Johnston}
  et~al.}{2020}]{Johnston2020}
{Johnston} S.,  et~al., 2020, \mn@doi [\mnras] {10.1093/mnras/staa516}, \href
  {https://ui.adsabs.harvard.edu/abs/2020MNRAS.493.3608J} {493, 3608}

\bibitem[\protect\citeauthoryear{{Kennea} et~al.,}{{Kennea}
  et~al.}{2013}]{Kennea2013}
{Kennea} J.~A.,  et~al., 2013, \mn@doi [\apjl] {10.1088/2041-8205/770/2/L24},
  \href {https://ui.adsabs.harvard.edu/abs/2013ApJ...770L..24K} {770, L24}

\bibitem[\protect\citeauthoryear{{Lang}, {Goss}  \& {Wood}}{{Lang}
  et~al.}{1997}]{Lang1997}
{Lang} C.~C.,  {Goss} W.~M.,   {Wood} O.~S.,  1997, \mn@doi [\apj]
  {10.1086/303452}, \href
  {https://ui.adsabs.harvard.edu/abs/1997ApJ...474..275L} {474, 275}

\bibitem[\protect\citeauthoryear{{Lazarian} \& {Pogosyan}}{{Lazarian} \&
  {Pogosyan}}{2016}]{Lazarian2016}
{Lazarian} A.,  {Pogosyan} D.,  2016, \mn@doi [The Astrophysical Journal]
  {10.3847/0004-637X/818/2/178}, \href
  {https://ui.adsabs.harvard.edu/abs/2016ApJ...818..178L} {818, 178}

\bibitem[\protect\citeauthoryear{{Levin} et~al.,}{{Levin}
  et~al.}{2012}]{Levin2012}
{Levin} L.,  et~al., 2012, \mn@doi [\mnras] {10.1111/j.1365-2966.2012.20807.x},
  \href {https://ui.adsabs.harvard.edu/abs/2012MNRAS.422.2489L} {422, 2489}

\bibitem[\protect\citeauthoryear{{Li}, {Bilous}, {Ransom}, {Main}  \&
  {Yang}}{{Li} et~al.}{2022}]{Li2022}
{Li} D.,  {Bilous} A.,  {Ransom} S.,  {Main} R.,   {Yang} Y.-P.,  2022, arXiv
  e-prints, \href {https://ui.adsabs.harvard.edu/abs/2022arXiv220507917L} {p.
  arXiv:2205.07917}

\bibitem[\protect\citeauthoryear{{Lorimer} \& {Kramer}}{{Lorimer} \&
  {Kramer}}{2004}]{Handbook}
{Lorimer} D.~R.,  {Kramer} M.,  2004, {Handbook of Pulsar Astronomy}.
 {Cambridge Observing Handbooks for Research Astronomers} Vol. 4, {Cambridge
  University Press}

\bibitem[\protect\citeauthoryear{{Lyne} \& {Lorimer}}{{Lyne} \&
  {Lorimer}}{1994}]{Lyne1994}
{Lyne} A.~G.,  {Lorimer} D.~R.,  1994, \mn@doi [\nat] {10.1038/369127a0}, \href
  {https://ui.adsabs.harvard.edu/abs/1994Natur.369..127L} {369, 127}

\bibitem[\protect\citeauthoryear{{Macquart}, {Ekers}, {Feain}  \&
  {Johnston-Hollitt}}{{Macquart} et~al.}{2012}]{Macquart2012}
{Macquart} J.~P.,  {Ekers} R.~D.,  {Feain} I.,   {Johnston-Hollitt} M.,  2012,
  \mn@doi [\apj] {10.1088/0004-637X/750/2/139}, \href
  {https://ui.adsabs.harvard.edu/abs/2012ApJ...750..139M} {750, 139}

\bibitem[\protect\citeauthoryear{{Martin}, {Walsh}, {Xiao}, {Lane}, {Walker}
  \& {Stark}}{{Martin} et~al.}{2004}]{Martin2004}
{Martin} C.~L.,  {Walsh} W.~M.,  {Xiao} K.,  {Lane} A.~P.,  {Walker} C.~K.,
  {Stark} A.~A.,  2004, \mn@doi [\apjs] {10.1086/379661}, \href
  {https://ui.adsabs.harvard.edu/abs/2004ApJS..150..239M} {150, 239}

\bibitem[\protect\citeauthoryear{{Mckinven} et~al.,}{{Mckinven}
  et~al.}{2022}]{Mckinven2022}
{Mckinven} R.,  et~al., 2022, arXiv e-prints, \href
  {https://ui.adsabs.harvard.edu/abs/2022arXiv220509221M} {p. arXiv:2205.09221}

\bibitem[\protect\citeauthoryear{{Minter} \& {Spangler}}{{Minter} \&
  {Spangler}}{1996}]{Minter1996}
{Minter} A.~H.,  {Spangler} S.~R.,  1996, \mn@doi [The Astrophysical Journal]
  {10.1086/176803}, \href
  {https://ui.adsabs.harvard.edu/abs/1996ApJ...458..194M} {458, 194}

\bibitem[\protect\citeauthoryear{{Mori} et~al.,}{{Mori}
  et~al.}{2013}]{Mori2013}
{Mori} K.,  et~al., 2013, \mn@doi [\apjl] {10.1088/2041-8205/770/2/L23}, \href
  {https://ui.adsabs.harvard.edu/abs/2013ApJ...770L..23M} {770, L23}

\bibitem[\protect\citeauthoryear{{Morris} \& {Serabyn}}{{Morris} \&
  {Serabyn}}{1996}]{Morris1996}
{Morris} M.,  {Serabyn} E.,  1996, \mn@doi [\araa]
  {10.1146/annurev.astro.34.1.645}, \href
  {https://ui.adsabs.harvard.edu/abs/1996ARA&A..34..645M} {34, 645}

\bibitem[\protect\citeauthoryear{{Noutsos}, {Johnston}, {Kramer}  \&
  {Karastergiou}}{{Noutsos} et~al.}{2008}]{Noutsos2008}
{Noutsos} A.,  {Johnston} S.,  {Kramer} M.,   {Karastergiou} A.,  2008, \mn@doi
  [MNRAS] {10.1111/j.1365-2966.2008.13188.x}, \href
  {https://ui.adsabs.harvard.edu/abs/2008MNRAS.386.1881N} {386, 1881}

\bibitem[\protect\citeauthoryear{{Oka}, {Geballe}, {Goto}, {Usuda}, {Benjamin},
  {McCall}  \& {Indriolo}}{{Oka} et~al.}{2019}]{Oka2019}
{Oka} T.,  {Geballe} T.~R.,  {Goto} M.,  {Usuda} T.,  {Benjamin} {McCall} J.,
  {Indriolo} N.,  2019, \mn@doi [\apj] {10.3847/1538-4357/ab3647}, \href
  {https://ui.adsabs.harvard.edu/abs/2019ApJ...883...54O} {883, 54}

\bibitem[\protect\citeauthoryear{{Orlando} \& {Strong}}{{Orlando} \&
  {Strong}}{2013}]{Orlando2013}
{Orlando} E.,  {Strong} A.,  2013, \mn@doi [\mnras] {10.1093/mnras/stt1718},
  \href {https://ui.adsabs.harvard.edu/abs/2013MNRAS.436.2127O} {436, 2127}

\bibitem[\protect\citeauthoryear{{Par{\'e}}, {Lang}, {Morris}, {Moore}  \&
  {Mao}}{{Par{\'e}} et~al.}{2019}]{Pare2019}
{Par{\'e}} D.~M.,  {Lang} C.~C.,  {Morris} M.~R.,  {Moore} H.,   {Mao} S.~A.,
  2019, \mn@doi [\apj] {10.3847/1538-4357/ab45ed}, \href
  {https://ui.adsabs.harvard.edu/abs/2019ApJ...884..170P} {884, 170}

\bibitem[\protect\citeauthoryear{{Par{\'e}}, {Purcell}, {Lang}, {Morris}  \&
  {Green}}{{Par{\'e}} et~al.}{2021}]{Pare2021}
{Par{\'e}} D.~M.,  {Purcell} C.~R.,  {Lang} C.~C.,  {Morris} M.~R.,   {Green}
  J.~A.,  2021, \mn@doi [\apj] {10.3847/1538-4357/ac2cc4}, \href
  {https://ui.adsabs.harvard.edu/abs/2021ApJ...923...82P} {923, 82}

\bibitem[\protect\citeauthoryear{{Pennucci} \& {Demorest}}{{Pennucci} \&
  {Demorest}}{2018}]{Pennucci2018}
{Pennucci} T.,  {Demorest} P.,  2018, {Pennucci/Pulseportraiture: First
  Official Release}, Zenodo, \mn@doi{10.5281/zenodo.1487794}

\bibitem[\protect\citeauthoryear{{Pennucci}, {Demorest}  \&
  {Ransom}}{{Pennucci} et~al.}{2016}]{Pennucci2016}
{Pennucci} T.~T.,  {Demorest} P.~B.,   {Ransom} S.~M.,  2016, {Pulse
  Portraiture: Pulsar timing}, Astrophysics Source Code Library, record
  ascl:1606.013 (\mn@eprint {ascl} {1606.013})

\bibitem[\protect\citeauthoryear{{Petroff}, {Hessels}  \& {Lorimer}}{{Petroff}
  et~al.}{2019}]{Petroff2019}
{Petroff} E.,  {Hessels} J.~W.~T.,   {Lorimer} D.~R.,  2019, \mn@doi [\aapr]
  {10.1007/s00159-019-0116-6}, \href
  {https://ui.adsabs.harvard.edu/abs/2019A&ARv..27....4P} {27, 4}

\bibitem[\protect\citeauthoryear{{Porayko} et~al.,}{{Porayko}
  et~al.}{2019}]{Porayko2019}
{Porayko} N.~K.,  et~al., 2019, \mn@doi [\mnras] {10.1093/mnras/sty3324}, \href
  {https://ui.adsabs.harvard.edu/abs/2019MNRAS.483.4100P} {483, 4100}

\bibitem[\protect\citeauthoryear{{Schnitzeler} \& {Lee}}{{Schnitzeler} \&
  {Lee}}{2015}]{Schnitzeler2015}
{Schnitzeler} D.~H.~F.~M.,  {Lee} K.~J.,  2015, \mn@doi [\mnras]
  {10.1093/mnrasl/slu171}, \href
  {https://ui.adsabs.harvard.edu/abs/2015MNRAS.447L..26S} {447, L26}

\bibitem[\protect\citeauthoryear{{Schnitzeler} \& {Lee}}{{Schnitzeler} \&
  {Lee}}{2017}]{Schnitzeler2017}
{Schnitzeler} D.~H.~F.~M.,  {Lee} K.~J.,  2017, \mn@doi [\mnras]
  {10.1093/mnras/stw3104}, \href
  {https://ui.adsabs.harvard.edu/abs/2017MNRAS.466..378S} {466, 378}

\bibitem[\protect\citeauthoryear{{Schnitzeler}, {Eatough}, {Ferri{\`e}re},
  {Kramer}, {Lee}, {Noutsos}  \& {Shannon}}{{Schnitzeler}
  et~al.}{2016}]{Schnitzeler2016}
{Schnitzeler} D.~H.~F.~M.,  {Eatough} R.~P.,  {Ferri{\`e}re} K.,  {Kramer} M.,
  {Lee} K.~J.,  {Noutsos} A.,   {Shannon} R.~M.,  2016, \mn@doi [\mnras]
  {10.1093/mnras/stw841}, \href
  {https://ui.adsabs.harvard.edu/abs/2016MNRAS.459.3005S} {459, 3005}

\bibitem[\protect\citeauthoryear{{Simpson}, {Colgan}, {Cotera}, {Erickson},
  {Hollenbach}, {Kaufman}  \& {Rubin}}{{Simpson} et~al.}{2007}]{Simpson2007}
{Simpson} J.~P.,  {Colgan} S. W.~J.,  {Cotera} A.~S.,  {Erickson} E.~F.,
  {Hollenbach} D.~J.,  {Kaufman} M.~J.,   {Rubin} R.~H.,  2007, \mn@doi [\apj]
  {10.1086/522295}, \href
  {https://ui.adsabs.harvard.edu/abs/2007ApJ...670.1115S} {670, 1115}

\bibitem[\protect\citeauthoryear{{Sobey} et~al.,}{{Sobey}
  et~al.}{2019}]{Sobey2019}
{Sobey} C.,  et~al., 2019, \mn@doi [\mnras] {10.1093/mnras/stz214}, \href
  {https://ui.adsabs.harvard.edu/abs/2019MNRAS.484.3646S} {484, 3646}

\bibitem[\protect\citeauthoryear{{Tendulkar} et~al.,}{{Tendulkar}
  et~al.}{2021}]{Tendulkar2021}
{Tendulkar} S.~P.,  et~al., 2021, \mn@doi [\apjl] {10.3847/2041-8213/abdb38},
  \href {https://ui.adsabs.harvard.edu/abs/2021ApJ...908L..12T} {908, L12}

\bibitem[\protect\citeauthoryear{{Tiburzi} et~al.,}{{Tiburzi}
  et~al.}{2013}]{Tiburzi2013}
{Tiburzi} C.,  et~al., 2013, \mn@doi [MNRAS] {10.1093/mnras/stt1834}, \href
  {https://ui.adsabs.harvard.edu/abs/2013MNRAS.436.3557T} {436, 3557}

\bibitem[\protect\citeauthoryear{{Verbunt}, {Igoshev}  \& {Cator}}{{Verbunt}
  et~al.}{2017}]{Verbunt2017}
{Verbunt} F.,  {Igoshev} A.,   {Cator} E.,  2017, \mn@doi [\aap]
  {10.1051/0004-6361/201731518}, \href
  {https://ui.adsabs.harvard.edu/abs/2017A&A...608A..57V} {608, A57}

\bibitem[\protect\citeauthoryear{{Wang}, {Han}  \& {Lai}}{{Wang}
  et~al.}{2011}]{Wang2011}
{Wang} C.,  {Han} J.~L.,   {Lai} D.,  2011, \mn@doi [\mnras]
  {10.1111/j.1365-2966.2011.19333.x}, \href
  {https://ui.adsabs.harvard.edu/abs/2011MNRAS.417.1183W} {417, 1183}

\bibitem[\protect\citeauthoryear{{Wang}, {Zhang}, {Dai}  \& {Cheng}}{{Wang}
  et~al.}{2022a}]{Wang2022}
{Wang} F.~Y.,  {Zhang} G.~Q.,  {Dai} Z.~G.,   {Cheng} K.~S.,  2022a, \mn@doi
  [Nature Communications] {10.1038/s41467-022-31923-y}, \href
  {https://ui.adsabs.harvard.edu/abs/2022NatCo..13.4382W} {13, 4382}

\bibitem[\protect\citeauthoryear{{Wang} et~al.,}{{Wang}
  et~al.}{2022b}]{Wang2022B}
{Wang} P.,  et~al., 2022b, The Astronomer's Telegram, \href
  {https://ui.adsabs.harvard.edu/abs/2022ATel15619....1W} {15619, 1}

\bibitem[\protect\citeauthoryear{{Yusef-Zadeh}, {Morris}  \&
  {Chance}}{{Yusef-Zadeh} et~al.}{1984}]{Yusef-zadeh1984}
{Yusef-Zadeh} F.,  {Morris} M.,   {Chance} D.,  1984, \mn@doi [\nat]
  {10.1038/310557a0}, \href
  {https://ui.adsabs.harvard.edu/abs/1984Natur.310..557Y} {310, 557}

\bibitem[\protect\citeauthoryear{{van Straten}, {Demorest}  \& {Oslowski}}{{van
  Straten} et~al.}{2012}]{vanStraten2012}
{van Straten} W.,  {Demorest} P.,   {Oslowski} S.,  2012, Astronomical Research
  and Technology, \href {https://ui.adsabs.harvard.edu/abs/2012AR&T....9..237V}
  {9, 237}

\makeatother
\end{thebibliography}




\appendix

\section{Details of RM fits}\label{rm_fits}

In Section \ref{sec:analysis} we described the procedure to determine the RM for each observation. Here we will provide details of the procedure and compare the results with those obtained from an RM synthesis code.

For each observation we perform flux and polarization calibration and we excise the frequency channels and time integrations that are affected by radio frequency interference (RFI) using packages from \texttt{PSRCHIVE}. After that we sum all of the integrations together, reduce the number of channels by 4 and create profiles of the pulsars with 1024 frequency channels maintaining all of the Stokes parameters. The number of channels was chosen to be 1024 in order to maximise the signal to noise (S/N) while keeping the intra-channel PA swing to a minimum. The maximum value of RM we can search using this number of channels is determined by the value that causes the PA between neighbouring channels (measured at the lowest frequency) to differ by more than $\pi$ radians and is $\sim 300,000$ \radm. This value is similar to the maximum RM value that can be searched in the data using RM synthesis \citep{Brentjens2005,Schnitzeler2015}. 

The next step is to select the phase window where pulsar signal is present. We also select a phase window where the signal is not present to estimate the properties of the baseline and remove it from the pulsar signal. This way we can make sure that only the signal that originates from the pulsar is analysed. For this reason and because the pulsar magnetosphere is not expected to contribute to Faraday rotation (eg. \citealt{Wang2011}), we expect the pulsars to be Faraday thin sources that present a single peak in the RM spectrum. 

\begin{figure*}
\centering
	\includegraphics[width=0.45\textwidth]{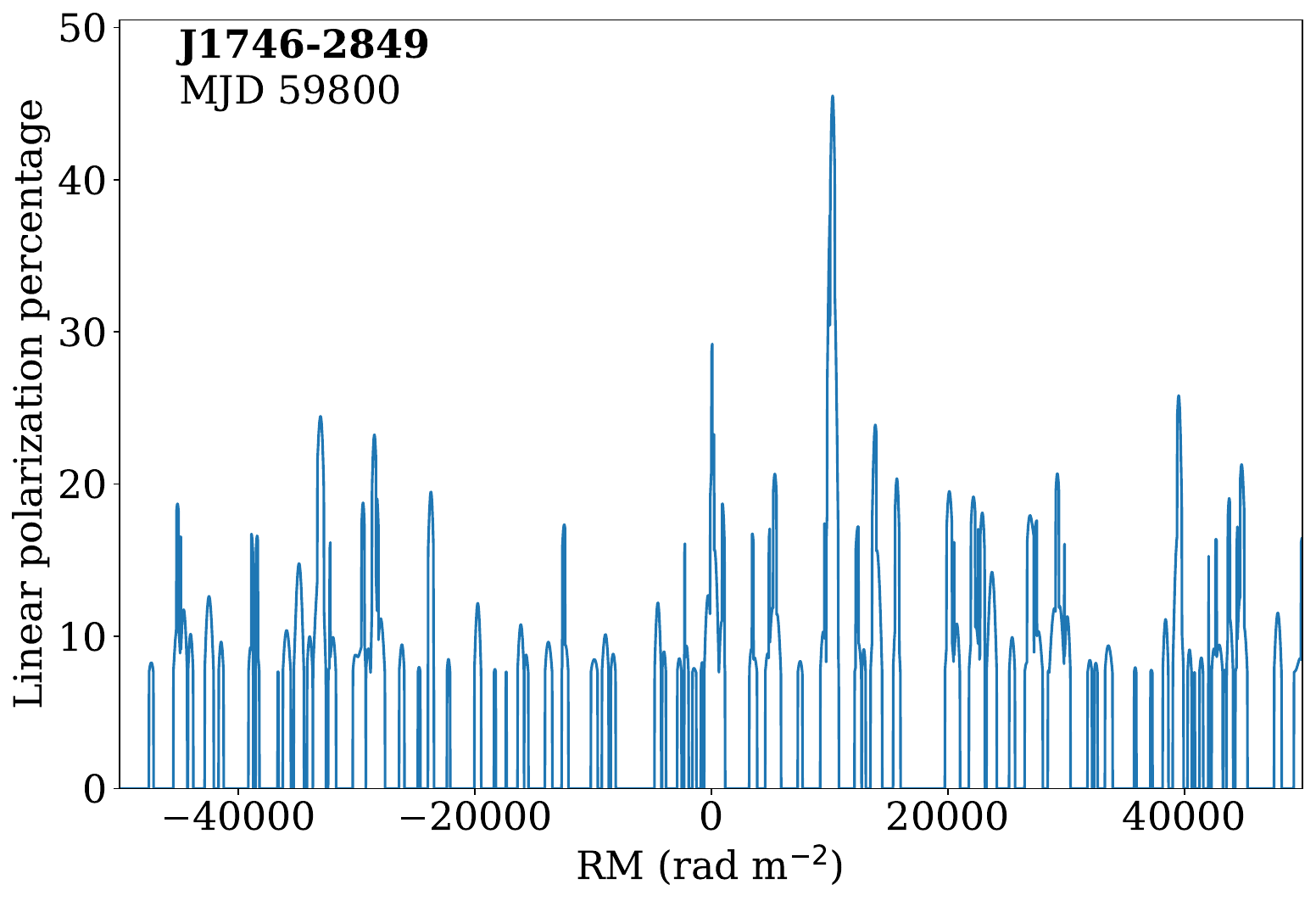}
	\,
	\includegraphics[width=0.45\textwidth]{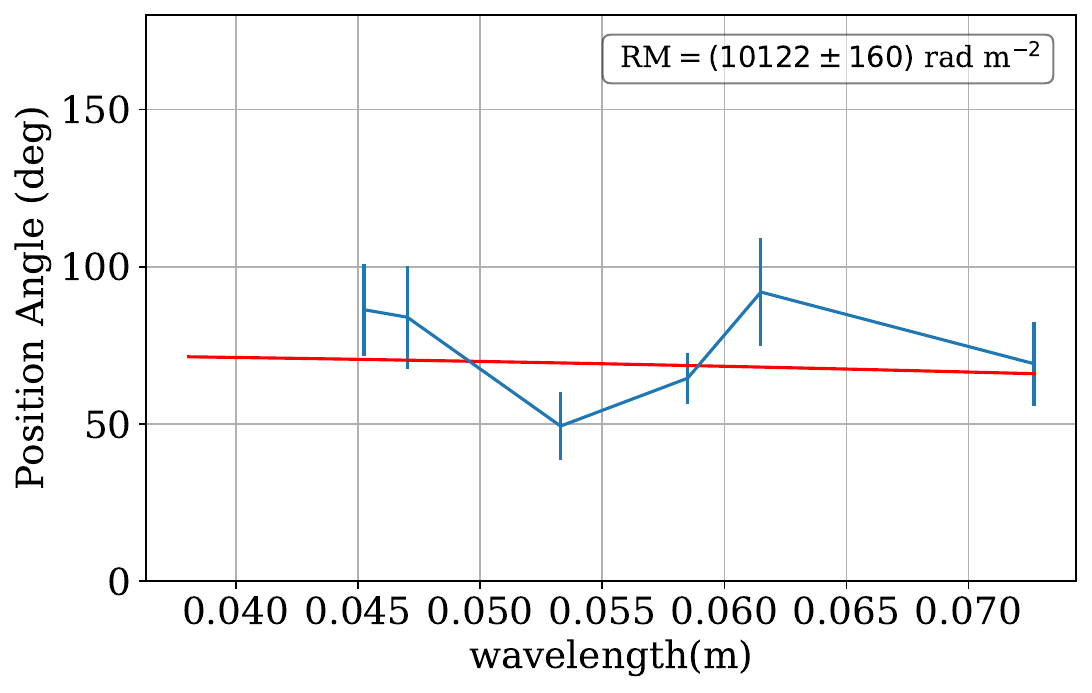}
	\,
	\includegraphics[width=0.45\textwidth]{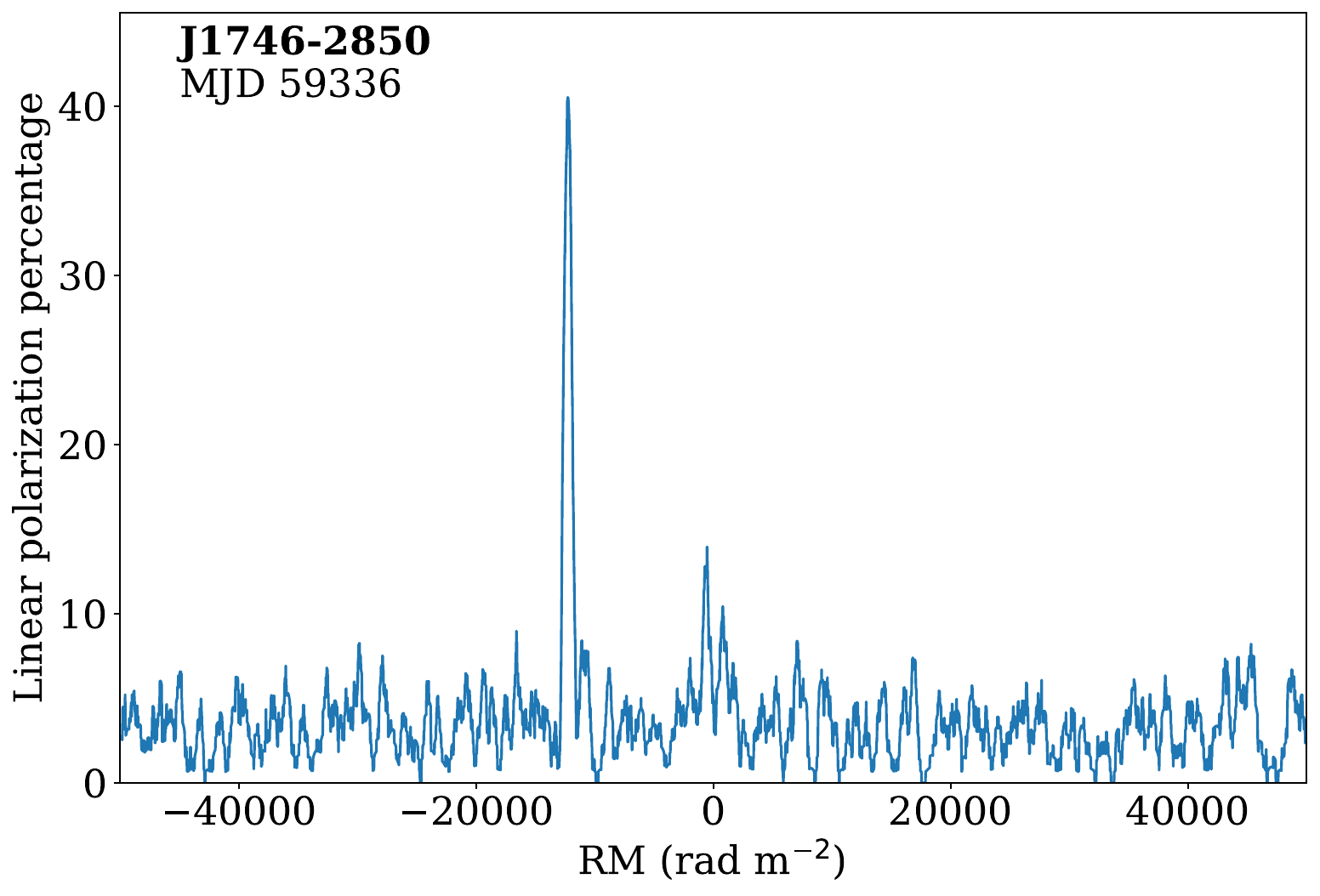}	
	\,
	\includegraphics[width=0.45\textwidth]{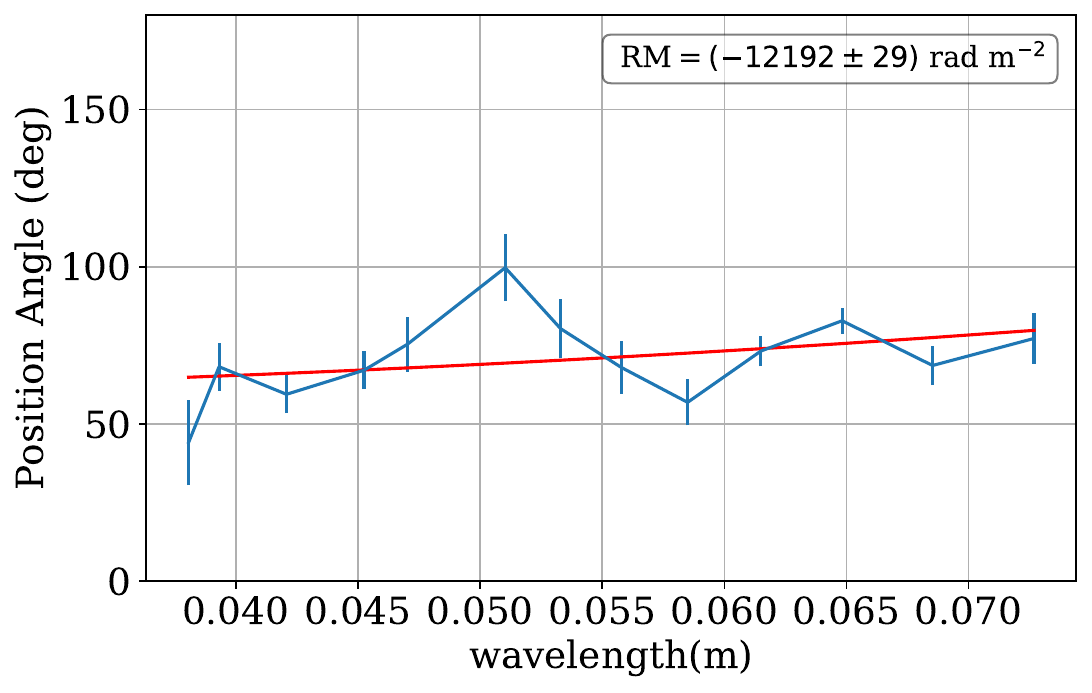}
 	\includegraphics[width=0.45\textwidth]{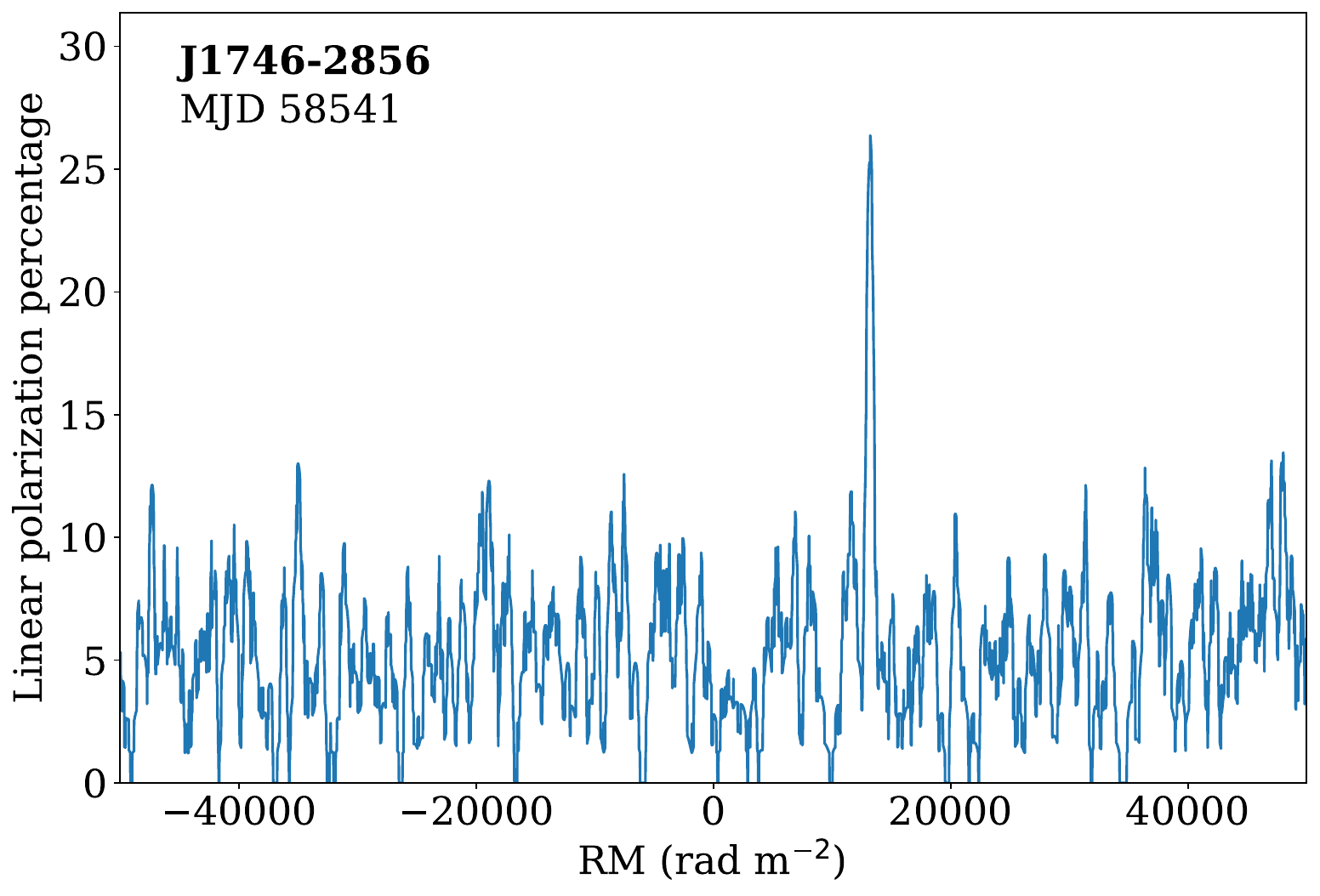}	
	\,
	\includegraphics[width=0.45\textwidth]{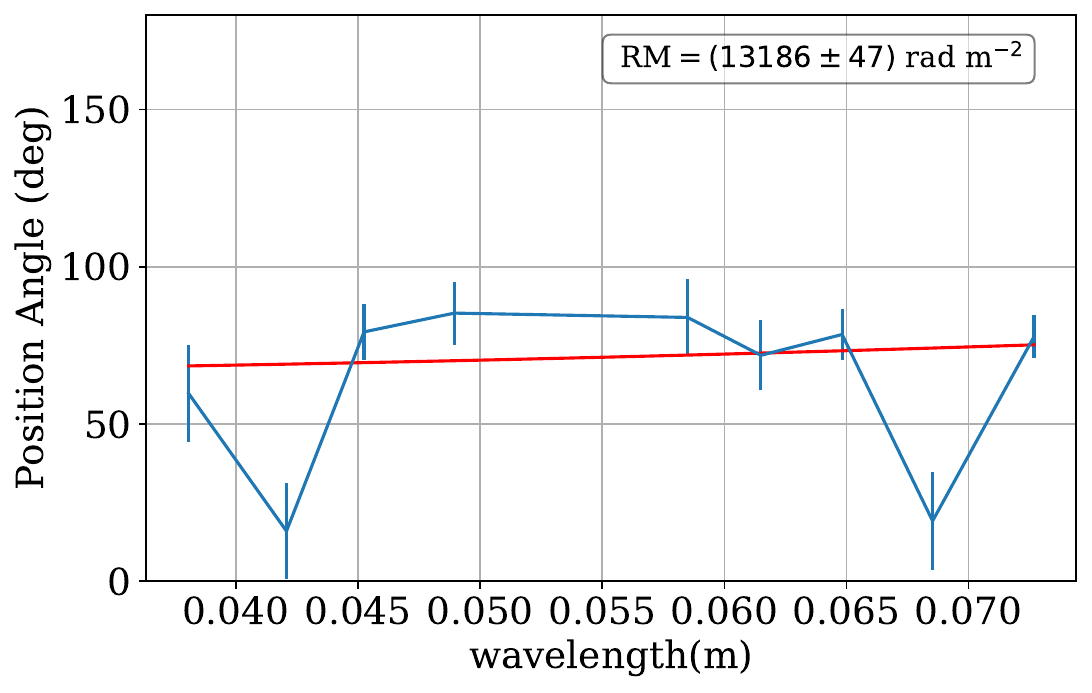}
 	\includegraphics[width=0.45\textwidth]{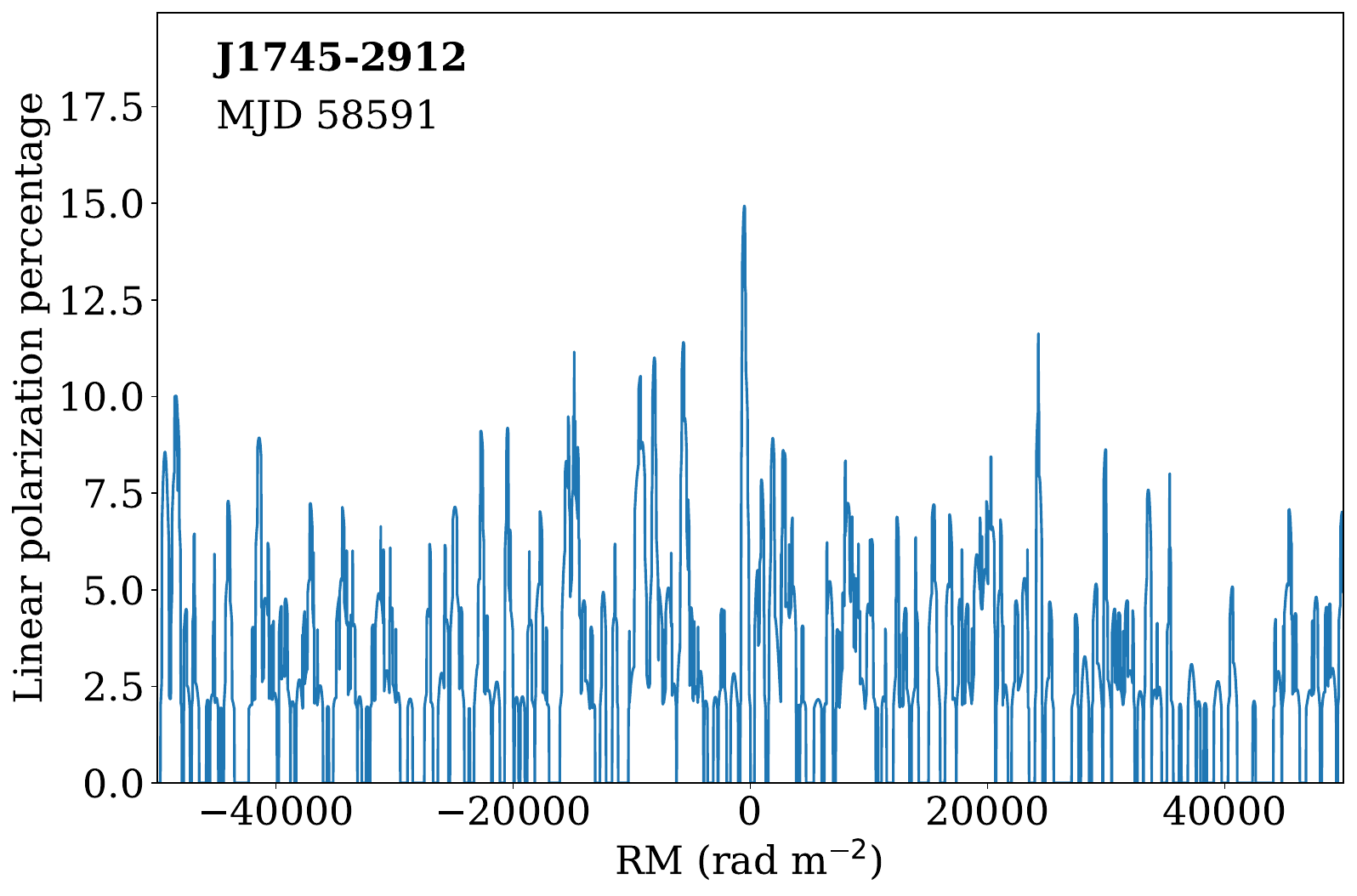}	
	\,
	\includegraphics[width=0.45\textwidth]{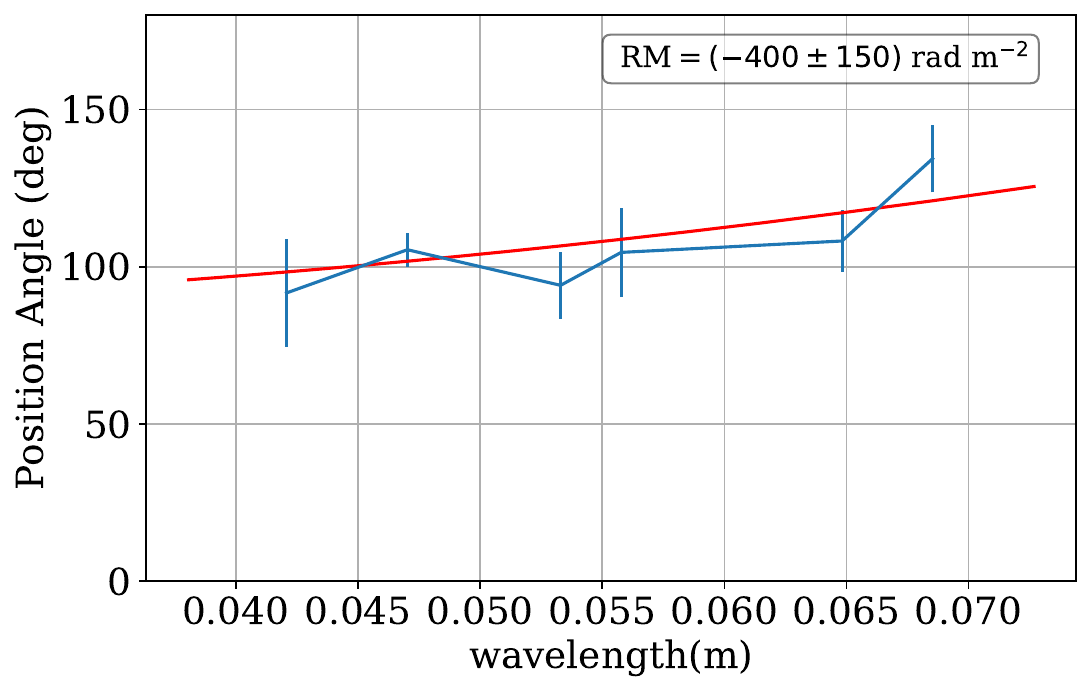}
  	\caption{ Examples of RM fits for one observation for each of the observed pulsars. In the left plot we show the linear polarization percentage as a function of the value of RM. The name of the pulsar and the day of the observation are reported in the top-left corner. In the right plot we show the fit of the position angle as a function of wavelength from the best-fitting value of RM. To avoid apparent phase wraps caused by the definition of the position angle, we plot it from 0 and 180 degrees. The best-fitting value of RM is shown in the top-right corner. For J1746-2849 and J1745-2912 we multiply the error on the RM by a factor of 2 since the S/N of the detection is smaller than 8. Further details of the plot are given in the text.
   }
  	\label{fig:RM_fits_examples}
\end{figure*}

The first method we used to measure the RM is to look for a peak in linear polarization in the RM spectrum of the pulsar. Given the range of RMs for the pulsars measured by \cite{Schnitzeler2016}, we decided to search for RMs in the range $-$50,000 - 50,000 \radm using steps of 10 \radm. Examples of these spectra for each of of the pulsars are shown in the left side of Figure \ref{fig:RM_fits_examples}. 
In order to check if the peak is significant enough to claim a detection, we perform a fit of the cumulative distribution function of the linear polarization percentage with a Gaussian function. We estimate the significance as the distance of the polarization peak from the mean value of the noise in number of standard deviations. 
We looked for an appropriate value of the threshold level by applying the same technique to noise taken from the off-pulse as input. In every case that we tested, the tallest peak in the RM spectrum for the noise has S/N $< 3.5$. Therefore we decided to use the value of S/N $>$ 4 for the threshold\footnote{ Given that the noise levels are slightly different for each observation and pulsar it is not straightforward to convert between this and the threshold levels suggested by \cite{Macquart2012,Hales2012}.}. The resulting S/N are reported in the Table \ref{tab:results_comparison} for the cases where the S/N is above the threshold.
A similar limit was used by \cite{Sobey2019} in their analysis of RM of pulsars. For observations with $4<$ S/N $<8$ we multiplied the error of the RM detection by a factor of 2 in order to consider the effects lower S/N as suggested by \cite{Sobey2019}. 

Once an initial value of RM for the observation has been found, we apply the correction to the profile, reduced the number of channels to 16 in order to increase the S/N ratio in each channel and proceed to perform a PA fit using formula \ref{eq:PAfit}. There are two reasons for correcting the observation for the RM value that maximises the linear polarization before attempting the fit: we increase the chance that the linear polarization in each channel is high enough to allow an accurate determination of the PA and we avoid the risk of multiple phase jumps occurring. A phase jump occurs when the difference in PA between the first and the last frequency channels is larger than 180 degrees. The value of RM necessary for a phase jump to occur along the entire frequency band is $750$ \radm. This means that if the difference between the value of RM used to correct the profile and real value is less than $750$ \radm we don't expect any phase jump to occur.
For the PA fit we followed the prescription described in \cite{Noutsos2008,Tiburzi2013,Abbate2020}. If the S/N ratio of the linear polarization in a channel is less than 2, the channel is ignored and the resulting PA is excluded from the fit.
Since the PA is periodic every 180 degrees, we allow for possible jumps of 180 degrees in the fit and show two realizations of the PA when necessary. The results are shown in the third column of Table \ref{tab:results_comparison} with the errors multiplied by two if the S/N of the polarization peak is between 4 and 8. 

\begin{figure*}
\centering
	\includegraphics[width=0.7\textwidth]{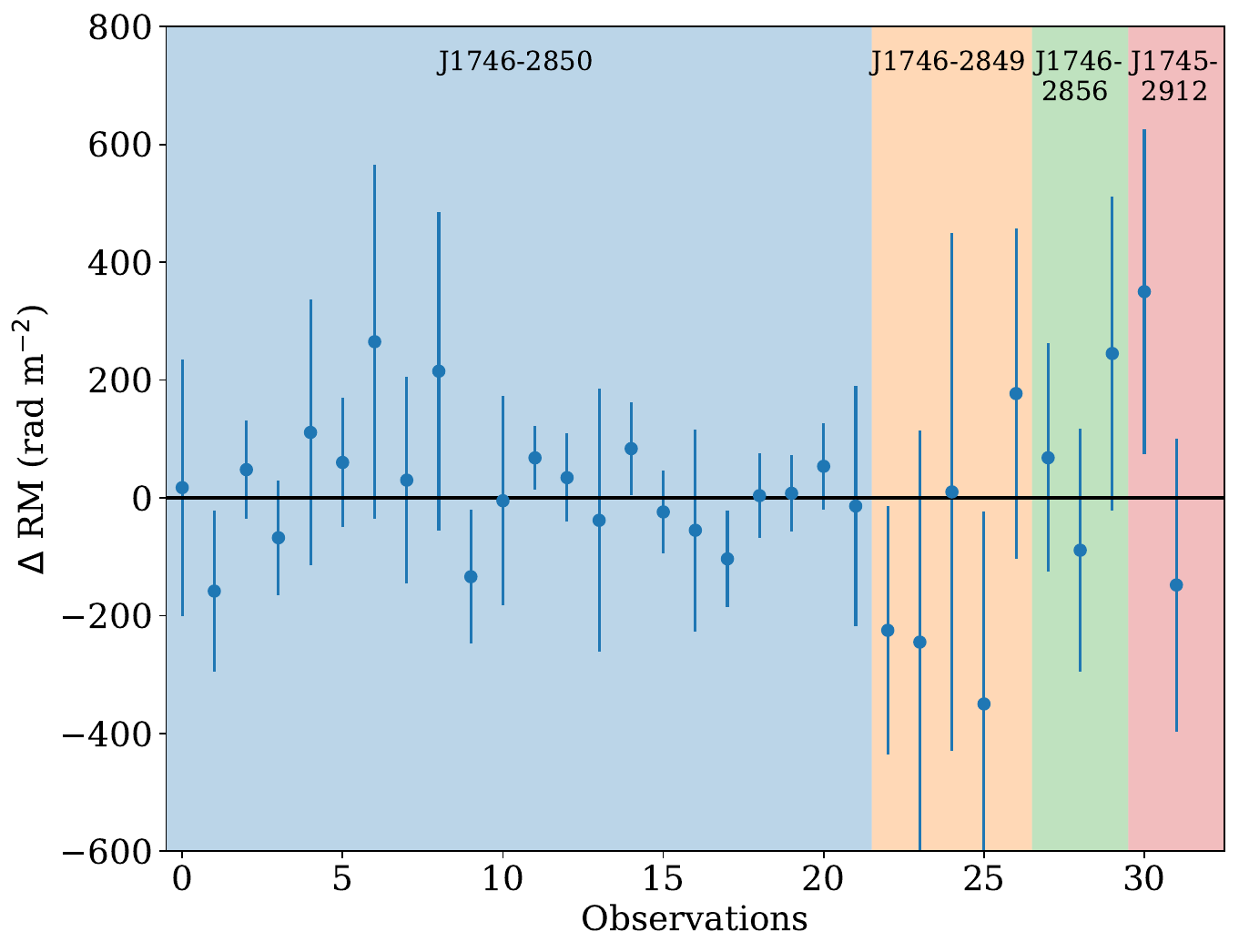}
    \caption{Plot of the differences between the RMs measured with the code described in the text and with RM synthesis. The colors indicate the different pulsars. The errors are shown at the $1\sigma$ level.
    }
  	\label{fig:RM_synthesis_comparison}
\end{figure*}

While similar methods have been used multiple times in the past \citep{Noutsos2008,Tiburzi2013,Han2018,Johnston2020}, it has some limitations. The cropping of channels with low S/N leads to a loss of part of the pulsar signal. Furthermore, the detection threshold and the error determination can be affected among other things by the spectral index of the pulsars \citep{Schnitzeler2017}. To test whether these limitations affect the reliability of the results, we repeated the analysis using the technique of RM synthesis \citep{Brentjens2005}. We used the code \texttt{RMcalc}\footnote{https://gitlab.mpifr-bonn.mpg.de/nporayko/RMcalc} \citep{Porayko2019} that implements RM synthesis. The results together with the S/N of the detection are shown in the forth and fifth column of Table \ref{tab:results_comparison}. Similarly to above we chose a threshold level of 4 to confirm a detection. There is one case for pulsar J1746-2849 where RM synthesis was able to detect the RM while our method wasn't and there is one case of the opposite happening for pulsar J1745-2912. For the cases where both methods returned a significant detection, the two values are compatible with one another at the $1\sigma$ level for most of them and at the $2\sigma$ level for the rest. The comparison is shown in Figure \ref{fig:RM_synthesis_comparison}.  
The successful comparison suggests that, for the observations presented in the paper, the method used is as accurate as RM synthesis.


\begin{table*}
\centering 
\caption{ Comparison between the RMs determined with the method described in the text and with RM synthesis. The column S/N (peak) show the significance of the peak of linear polarization. While the column S/N (syn) is the significance of the RM synthesis detection. If the significance in is less than 4, the corresponding value is ignored. The errors in the brackets are at $1\sigma$ level.}
\label{tab:results_comparison}
\footnotesize
\renewcommand{\arraystretch}{1}
\vskip 0.1cm
\begin{tabular}{c|ccccc}
\hline
Pulsar name  & Date (MJD) & \multicolumn{1}{c}{RM} & S/N (peak)& \multicolumn{1}{c}{RM synthesis} & S/N (syn)\\
 & &  \multicolumn{1}{c}{(rad m$^{-2}$)} &  & \multicolumn{1}{c}{(rad m$^{-2}$)}& \\
\hline
PSR J1746$-$2849 & 58544 & 9980(160) & 6.7 &  10205(51) & 8.1\\
           & 58557 &  9800(160)   & 4.6 & 10045(200) & 4.1 \\
           & 58623 &             &     & 9940(130) & 6.4 \\
           & 58865 &  10050(260) & 4.5 & 10040(180) & 4.8 \\
           & 59400 &  9920(146)   & 5.7 & 10270(180)  & 4.7 \\
           & 59480 &             &     &            &  \\
           & 59800 &  10122(160)  & 4.7 & 9945(120) & 7.0 \\
\hline           
PSR J1746$-$2850 & 58544 & -12363(44) & 9.4 & -12380(174) & 4.7\\
           & 58557 & -12533(86) &  7.3 & -12375(50) &  8.2\\
           & 58591 & -12507(33) &  9.8 & -12555(50) &  8.2\\
           & 58600 & -12453(68) & 11.5 & -12385(30) & 14.3\\
           & 58610 & -12529(46) & 9.3 & -12640(180) &  4.7\\
           & 58623 & -12510(60) & 11.3 & -12570(50) &  9.0\\
           & 58764 & -12340(120)& 4.2 & -12605(180) &  4.7 \\
           & 58817 & -12335(50) & 13.1 & -12365(128) &  6.7 \\
           & 58831 & -12130(80) & 9.0 & -12345(190) &  4.4\\
           & 58865 & -12504(64) & 8.2 & -12370(50) &  8.0\\
           & 59314 & -12185(34) & 11.9 & -12180(144) &  5.9\\
           & 59336 & -12192(29) &  21.2 & -12260(25) & 16.2\\
           & 59357 & -12006(31) &  20.5 & -12040(44) &  9.3\\
           & 59377 & -12368(43) &  6.5 & -12330(138) &  5.9\\
           & 59437 & -12186(32) &  22.7 & -12270(47) &  8.6 \\
           & 59473 & -12199(23) & 18.4 & -12175(47) &  8.8\\
           & 59502 & -12240(33) &  14.9 & -12185(138) &  5.9\\
           & 59545 & -12228(40) &  14.4 & -12125(42) &  9.8\\
           & 59577 & -12166(29) & 21.3 & -12170(43) &  9.5\\
           & 59608 & -12167(34) &  18.2 & -12175(30) & 13.9\\
           & 59629 & -12116(29) &  20.7 & -12170(44) &  9.5\\
           & 59800 & -12034(100)& 8.6 & -12020(208) &  4.0 \\
\hline
PSR J1746$-$2856 & 58541 & 13186(47) & 8.3 & 13295(152) & 5.4\\
           & 58610 &  13223(50) & 9.4 & 13155(144) &5.7\\
           & 58831 &  13096(100)& 7.9 & 13185(106) &7.8\\
\hline
PSR J1745$-$2912 & 58541 &  &  & & \\
           & 58591 & -400(150)& 4.0 & -645(192) & 4.3\\
           & 58801 & -480(146)& 4.1 & & \\
           & 58817 & -648(144)& 4.4 & -500(210) & 4.0 \\
\hline
\end{tabular}
\end{table*}

\bsp	
\label{lastpage}
\end{document}